\documentclass[fleqn,usenatbib,useAMS]{mnras}

\usepackage{newtxtext,newtxmath}

\usepackage[T1]{fontenc}
\usepackage{ae,aecompl}
\usepackage{xcolor}

\usepackage{graphicx}   
\usepackage{amsmath}    
\usepackage{amssymb}    
\usepackage{relsize}    

\newlength{\VSpaceBeforeTabBib}
\newlength{\VSpaceBeforeTabFoot}
\usepackage{xcolor}
\usepackage{xifthen}
\usepackage[normalem]{ulem}
\newcommand{\stkout}[1]{\ifmmode\text{\sout{\ensuremath{#1}}}\else\sout{#1}\fi}
\newcommand{\edited}[2]{\ifthenelse{\isempty{#1}}{\textcolor{red}{#2}}{\ifthenelse{\isempty{#2}}{\textcolor{gray}{\stkout{#1}}}{\textcolor{gray}{\stkout{#1}} \textcolor{red}{#2}}}}

\title{Grain Growth During Protostellar Disk Formation} 
\author[Y. Tu et al.]{
Yisheng Tu$^{1}$\thanks{yt2cr@virginia.edu}, Zhi-Yun Li$^{1}$, Ka Ho Lam$^{1}$
\\
$^{1}$Department of Astronomy, University of Virginia, Charlottesville, VA 22904, USA\\
}

\date{Accepted XXX. Received YYY; in original form ZZZ}

\pubyear{2019}

\begin{document}
\label{firstpage}
\pagerange{\pageref{firstpage}--\pageref{lastpage}}
\maketitle

\begin{abstract}

 Recent observations indicate that mm/cm-sized grains may exist in the embedded protostellar disks. How such large grains grow from the micron size (or less) in the earliest phase of star formation remains relatively unexplored. In this study we take a first step to model the grain growth in the protostellar environment, using two-dimensional (2D axisymmetric) radiation hydrodynamic and grain growth simulations. We show that the grain growth calculations can be greatly simplified by the ``terminal velocity approximation", where the dust drift velocity relative to the gas is proportional to its stopping time, which is proportional to the grain size. We find that the grain-grain collision from size-dependent terminal velocity alone is too slow to convert a significant fraction of the initially micron-sized grains into mm/cm sizes during the deeply embedded Class 0 phase. Substantial grain growth is achieved when the grain-grain collision speed is enhanced by a factor of 4. The dust growth above and below the disk midplane enables the grains to settle faster towards the midplane, which increases the local dust-to-gas ratio, which, in turn, speeds up further growth there. How this needed enhancement can be achieved is unclear, although turbulence is a strong possibility that deserves further exploration.  
\end{abstract}




\section{Introduction} 
\label{sec:intro}

The formation of a stellar system can be categorized generally into four classes based on their observational properties such as their spectral energy distribution \citep{Lada1987, Andre1993}. These observational differences are believed to be associated with their stages of evolution in the process of star (and planet) formation. In Class 0 sources, the protostar is deeply embedded in the surrounding envelope \citep{Chandler1998}, and during the evolution from Class 0 to Class I, the mass in the envelope is transferred to the protostar and protostellar disk. The system depletes its envelope in the Class II phase, forming a protoplanetary disk. The disk is eventually depleted in the Class III phase, leaving behind planets orbiting a stellar system. 

Planets are formed in circumstellar disks, but exactly when and how their formation starts in the disk remains uncertain.  Recent high-resolution surveys of Class II (protoplanetary) disks show substructures that potentially indicate the presence of planets \citep{Zhang_2016, Isella2016, Fedele2018, Andrews2018, Andrews2020}. Such substructures have now been observed in the younger Class I \citep{Sheehan2018, Segura-Cox2020} and even Class 0 disks \citep{Sheehan2020}. Some of the putative planets have inferred masses of the Jupiter or even larger, and may have reached the end of their growth judging from their low accretion rates \citep{Haffert2019, CrIdland2021}.

These observational evidences suggest that at least some planets must have formed early in the process of star formation. In order to form planets, dust grains in the protostellar disk must have grown from their sizes in the interstellar medium (ISM), typically of order $\sim0.1~\mu$m \citep{Mathis_et_al.1977}, to sizes that are large enough to partially decouple from the gas and potentially trigger streaming instability to start the formation of planetesimals and ultimately planets. This process demands the formation of a substantial amount of mm/cm-sized grains during the protostellar phase. 

There are evidences for large mm/cm sized grains in protostellar sources based on the spectral energy distribution \citep{Testi2014review}. Specifically, a value of the opacity index $\beta$ in the millimeter regime ($=\alpha-2$ in the optically thin limit, where $\alpha$ is the spectral index) of order 1 or less is usually interpreted as evidence for mm/cm-sized grains \citep{Draine2006, Beckwith1999}. For example, \citet{Galametz2019} found 7 out of 10 observed Class 0 sources exhibit $\beta_\mathrm{1-3mm} < 1$ within $1 \sigma$ after accounting for free-free emission coming from the central protostar. Other observational studies such as \citet{Jorgensen2007}, \citet{Tobin2013} and \citet{Kwon2015} have shown additional examples of protostellar sources exhibiting $\beta < 1$, indicating that mm/cm-sized grains are prevalent in the earliest phases of star formation\footnote{ There are, however, counter examples where the spatially resolved opacity index $\beta$ is close to the ISM value that is indicative of small grains (e.g., \citet{Ohashi2022} for the L1527 protostellar disk).}.

How the grains grow from (sub)$\mu$m sizes to mm/cm sizes in the protostellar phase is unclear. 
An early attempt to address this question was made by \citet{Morfill1978}, who performed a dust evolution calculation on a coarse 2D (axisymmetric) grid with an underlying analytic 1D (spherical) gas dynamic model. They found early grain growth to sizes up to $\sim 100\mu$m in the collapsing envelope around the protostar on a time scale of $1.67$~Myr, and proposed the idea that fast grain growth during the protostellar collapse may form seeds to later planet formation.  
\citet{Birnstiel2010} improved upon \citet{Morfill1978} by including both the collapsing protostellar envelope and a growing protostellar disk, as well as new physical ingredients for grain growth, notably, turbulence-enhanced grain-grain collision speeds \citep[see e.g.][]{Voelk1980, MMV1991, OrmelCuzzi2007}. However, their analytical treatment of the gas dynamics is essentially one dimensional (in the cylindrically radial direction), which leaves open the question of how the dust and gas enter the growing disk from the collapsing envelope from all directions and how the grains move and grow inside the disk as they gradually settle towards the midplane gravitationally. \citet{Suttner2001} performed a series of 2D (axisymmetry) models, but their calculations focused primarily on the effects of relatively moderate grain growth (to several tens of microns) on the optical/IR appearance of disks through its impact on dust opacities. \citet{Hirashita&Li2013} examined the grain growth in dense molecular cloud cores and concluded that the core lifetime must be several times the free-fall time to grow grains from the classical ISM sizes to micron-sizes \citep[see also][]{Ormel2009}. \citet{Wong2016} extended \citet{Hirashita&Li2013} to show that the dense molecular cores have too low densities for grains to grow to mm-sizes effectively. {More recently, \citet{Ohashi2021} showed the existence of dust ring structures around Class 0/I objects using analytical arguments assuming the dust growth time scale is comparable to the disk age, which are consistent with the ALMA observations. The existence of a dust ring also agrees with the need for the formation of pebble drifting in inner disks to allow pebble accretion, which can grow planets more effectively than the accretion of large planetsimals \citep[][]{Kobayashi2021}. \citet{Tsukamoto2021} proposed the idea of increased grain growth rate by circulating large grains in disks via ``ash-fall", in which large grains in the inner disk are carried out of the disk by magnetized outflows and re-enter the disk at large radii.}

%

In this work, we aim to investigate the dust dynamics and growth during the protostellar phase of the star and disk formation in two dimensions (2D) using a radiation hydrodynamical code, under the assumption of axisymmetry. We focus on grain growth to mm/cm sizes that are important for (sub)millimeter and centimeter dust continuum observations and planet formation. 
In section \S~\ref{sec:setup} we describe the hydro and dust simulation methods. 
In section \S~\ref{sec:hydro_results} and section \S~\ref{sec:dust_result} we present the results of the hydro and dust models respectively. We find that it is difficult to produce a significant amount of large mm/cm-sized grains during the protostellar phase in the baseline case where grains collide with one another only through the size-dependent dust-gas drift velocities (\S~\ref{sec:dust_the_fiducial_model}) and that significant growth can be achieved if the grain-grain collision speed is enhanced by a factor of 4 (\S~\ref{sec:dust_the_chi=4_model}). We obtain a simple expression for the grain size $e$-folding time in section \S~\ref{sec:sol_to_smoluchowski}, which is used to understand the numerical results on grain growth in section \S~\ref{sec:discussion}. We conclude in section \S~\ref{sec:conclusion}. 

\section{Problem Setup}
\label{sec:setup}

\subsection{Governing Equations for Gas Dynamics}
\label{sec:equations}

Protostellar disk formation in dense cores of molecular clouds is a complex process involving turbulence and magnetic fields \citep[and associated non-ideal MHD effects; see][for a review]{Li2014}. As a first step in exploring how the grains grow as the infalling envelope is transformed into a rotationally supported disk, we have decided to limit our investigation to the simplest case of the collapse of a rotating core without any magnetic field or turbulence.
The dynamics of core collapse and disk formation are governed by the following radiation hydrodynamical equations:
\begin{equation}
  \frac{\partial \rho}{\partial t} + \nabla \cdot \left( \rho \mathbfit{u} \right) = 0,
\end{equation}
\begin{equation}
  \label{eq:momentum}
  \rho \frac{\partial \mathbfit{u}}{\partial t} + \rho \left( \mathbfit{u} \cdot \nabla \right) \mathbfit{u} = -\nabla P - \rho \mathbfit{g} + \nabla\cdot \mathbf{\Pi} - \mathbfit{G}_r, 
\end{equation}
\begin{equation}
    \frac{\partial E}{\partial t} + \nabla\cdot[(E+P)\mathbfit{u} + \mathbf{\Pi}\cdot\mathbfit{v}] = \rho\mathbfit{u}\cdot\mathbfit{g} - G_r^0,
\end{equation}
\begin{equation}
    \frac{\partial \mathbfit{I}}{\partial t} + c\mathbfit{n}\cdot\nabla \mathbfit{I} = \mathbfit{S}(\mathbfit{I}, \mathbfit{n}),
    \label{equ:rt}
\end{equation}
where $\rho, \mathbfit{u}, P, E$ and $I$ are gas density, velocity, gas pressure, total energy density and frequency-integrated specific intensity respectively. $\mathbfit{g}$ is the gravitational acceleration, including both gas-self gravity and the gravity due to the central protostar. $\mathbfit{G}_r$ and $G_r^0$ are radiative force and the radiative heating/cooling rates respectively. The last equation, equation~(\ref{equ:rt}), is the standard time-dependent radiative transfer equation for the specific intensity $I$ along direction $\mathbfit{n}$, with $S$ denoting the source term (representing additional cosmic-ray heating).

The quantity $\mathbf{\Pi}$ is the viscous stress tensor given \textbf{by}
\begin{equation}
    \Pi_{ij} = \rho\nu\Big(\frac{\partial u_i}{\partial x_j} + \frac{\partial u_j}{\partial x_i} - \frac{2}{3}\delta_{ij}\nabla\cdot\mathbfit{u}\Big)
\end{equation}
where $\nu$ is the the kinematic viscosity. $\mathbf{\Pi}$ is used to model angular momentum transport in the absence of a magnetic field. We adopt a standard $\beta$-parametrization for the kinematic viscosity $\nu$ \citep[see e.g.][]{Kuiper_et_al._2010}
\begin{equation}
    \nu = \beta_\nu \Omega_K(r) R^2, 
    \label{eq:viscosity}
\end{equation}
where $R$ is the cylindrical radius and $\Omega_K(r)=\sqrt{GM(r)/r^3}$, with $M(r)$ denoting the mass enclosed within a sphere of radius $r$. For a rotationally supported disk, the quantity $\Omega_K$ would be the orbital angular velocity and the dimensionless parameter $\beta_\nu$ would be related to the standard $\alpha-$viscosity parameter by $\beta_\nu=\alpha (H/R)^2$ where $H$ is the disk scale height. 
%
%

\subsection{Hydrodynamic Numerical Method}

The governing equations are solved with the Athena++ code \citep[][]{Stone2020} under the assumption of (2D) axisymmetry around the rotation axis in a spherical polar coordinate system $(r,\theta,\phi)$. We adopt a computation grid with logarithmic spacing in the radial direction and constant spacing in the $\theta$-direction. Logarithmic spacing has the advantage of providing higher resolution at smaller  radii relevant to disk formation, at the expense of producing larger cells at larger radii, which are not as critical for our analysis. The grid extends logarithmically in the radial direction from $10$ to $10^4$~au with 320 cells, and linearly in the $\theta$-direction from $0$ to $\pi$ with 208 cells.

There are 4 boundary conditions treated in the simulation. The inner and outer radial boundaries are treated \textbf{in} the same way, with a semi-outflow boundary condition where material is allowed to exit the computational domain but not enter back in. The standard reflective boundary conditions are imposed at the poles.  

The simulation is initialized with an uniform temperature 8K and a power-law density profile following \citet{Shu1977}
\begin{equation}
    \rho_\mathrm{g}(r, t = 0) = \frac{A_sc_s^2}{4\pi G r^2}
\end{equation}
where $c_s = \sqrt{\frac{k_b T}{\mu m_H}}$ is the isothermal sound speed and $A_s$ is chosen to be $3.11$. This yields $\sim 1.66 M_\odot$ inside the simulation domain. The central protostar has a small initial mass of $0.0017 M_\odot$, which is increased as mass is accreted across the inner radial boundary. Effects of stellar radiation are included using the  stellar evolution model of \citet{Hosokawa&Omukai2009} that assumes an accretion rate $10^{-6}M_\odot\ \mathrm{yr}^{-1}$. It yields a stellar luminosity $L_*$ and a stellar radius $R_*$ for a given stellar mass $M_*$. The total luminosity is given by 
\begin{equation}
    L = L_* + L_\mathrm{acc} = L_* +\frac{ G\ M_*\ \dot{M}_*}{R_*}
    \label{equ:tot_luminosity}
\end{equation}
where the accretion rate $\dot{M}_*$ is found by measuring the amount of mass going through inner boundary in our simulation. The total luminosity is then distributed uniformly onto the inner boundary of our simulation at $r = 10$~au for the radiative transfer calculation by the code. 

For the radiative transfer in the simulation domain, we adopted an opacity $\kappa_\nu$ model from \citet{Laor&Draine1993} and \citet{Kuiper_et_al._2010}, with a constant dust-to-gas ratio of 0.01. 
To speed up the simulation, we reduced the speed of light by a factor of $10^4$ \citep[see e.g.][]{Chang_et_al.2020}.

We assume $\phi$ displacement symmetry but allow for a $\phi$ direction velocity. The initial $\phi$ direction velocity is determined by the solid-body rotation rate $\Omega_\mathrm{init}$, taken to be $3.75$ rad/Myr. Viscosity is also included using the model described by equation~\ref{eq:viscosity}, with a constant $\beta_\nu = 0.001$. This combination of parameters yields a hundred au-sized disk towards the end of the simulation.

\subsection{Modeling Grain Dynamics and Growth}
\label{sec:grain setup}

We post-process the hydrodynamical outputs from Athena++ for grain dynamics and coagulation calculations. The hydrodynamical results are outputted with 50 years intervals so that the hydrodynamical evolution can be reconstructed with linear interpolation between consecutive frames. In this study, we consider spherical dust particles, with radius determined by the mass $m$ of the particle
\begin{equation}
    s = \Big(\frac{3\ m}{4\pi\tilde{\rho}_\mathrm{dm}}\Big)^{1/3}    
\end{equation}
where $\tilde{\rho}_\mathrm{dm}$ is the material density of the grain. 

The grains are evolved on a grid with the same resolution as the hydrodynamical grid, but only include the upper hemisphere by taking advantage of the disk being symmetric about the midplane. Reflective boundary conditions are applied on the midplane. The grain sizes are discretized into 105 bins logarithmic spaced from $0.1\,\mu\mathrm{m}$ to $1\,\mathrm{cm}$. The number of size bins is chosen to be as large as possible to minimize numerical diffusion into large grains \citep[see e.g.][]{Lombart2021}, while keeping a reasonable simulation run-time. In all models, the grain everywhere is initialized from $0.1\,\mu\mathrm{m}$ to $1\,\mu\mathrm{m}$ following an MRN-distribution\footnote{Our choice of an initial maximum grain size of $1\,\mu\mathrm{m}$ is motivated by the detection of mid-infrared emission in dense cores of molecular clouds produced via scattering by micron-sized grains \citep{Steinacker_et_al.2010}} $n(s)\propto s^{-3.5}$ \citep{Mathis_et_al.1977}, assuming an initial dust-to-gas mass ratio 0.01. We evolve the number density in each grain size bin ($N_i$) in our numerical model, which is related to the usual number density per unit size $n(s)$ through
\begin{equation}
    N_i = \int_{s_i}^{s_{i+1}} n(s) ds
    \label{equ:defineN_ibyn_i}
\end{equation}
where $s_i$ and $s_{i+1}$ are the grain sizes in two adjacent size bins. 

The grain distribution on the mesh grid is initialized approximately 2500 years before the appearance of a disk. Because the high density region of the envelope falls quickly through our inner boundary prior to disk formation and the remaining envelope is orders of magnitude less dense than the majority of the disk, grains are not expected to grow significantly before the formation of the disk because of the low density and short core collapse time (see section \S~\ref{sec:sol_to_smoluchowski}).  

\subsubsection{Grain dynamics}

The acceleration of grains of different sizes can be modeled with the pressure-less non-viscous Navier-Stokes equations in spherical polar coordinate, given by \citep{acheson1990elementary}
\begin{equation}
    \begin{split}
        &\frac{\partial u_r}{\partial t} = F_r  - u_r\frac{\partial u_r}{\partial r} - \frac{u_\phi}{r\sin\theta}\frac{\partial u_r}{\partial\phi} - \frac{u_\theta}{r}\frac{\partial u_r}{\partial\theta} + \frac{u_\theta^2 + u_\phi^2}{r}\\
        &\frac{\partial u_\theta}{\partial t} = F_\theta - u_r\frac{\partial u_\theta}{\partial r} - \frac{u_\phi}{r\sin\theta}\frac{\partial u_\theta}{\partial\phi} - \frac{u_\theta}{r}\frac{\partial u_\theta}{\partial\theta} - \frac{u_ru_\theta - u_\phi^2\cot\theta}{r} \\
        &\frac{\partial u_\phi}{\partial t} = F_\phi - u_r\frac{\partial u_\phi}{\partial r} - \frac{u_\phi}{r\sin\theta}\frac{\partial u_\phi}{\partial\phi} -\frac{u_\theta}{r}\frac{\partial u_\phi}{\partial\theta} - \frac{u_ru_\phi + u_\phi u_\theta\cot\theta}{r}
    \end{split}
\end{equation}
where $F_r$, $F_\theta$ and $F_\phi$ denotes the body force per unit mass acting on the dust grains, including both gravity and hydrodynamical drag. $u_r$, $u_\theta$ and $u_\phi$ are the dust velocities in $r, \theta$ and $\phi$ directions respectively. In the remainder of this subsection, we will use these notations to represent dust grain speeds.
Because we are interested in relatively small grains ($0.1\,\mu\mathrm{m}\to1\,\mathrm{cm}$), the drag force is in the Epstein regime, where the stopping time $t_s$ is given by \citep{Epstein1924, Armitage2015}
\begin{equation}
    t_s = \frac{\tilde{\rho}_\mathrm{dm} s}{\rho_\mathrm{g} v_\mathrm{th}}
    \label{equ:stopping_time}
\end{equation}
where $v_\mathrm{th} = \sqrt{\frac{8k_bT}{\pi \mu m_H}}$ is the gas thermal velocity, $k_b$ and $T$ are the Boltzmann constant and temperature respectively, $\mu = 2.33$ and $m_H$ are the mean particle weight and hydrogen mass, and  $\rho_\mathrm{g}$ is the gas density at the locations of the dust grains. The force per unit mass $\mathbfit{F}$ is given by
\begin{equation}
    \mathbfit{F} = \rho\mathbfit{g} + \rho\frac{\mathbfit{u}_\mathrm{gas} - \mathbfit{u}_\mathrm{dust}}{t_s}
    \label{equ:equation_of_motion}
\end{equation}
where $\mathbfit{g}$ is the local gravitaional acceleration and $\mathbfit{u}_\mathrm{gas}$ and $\mathbfit{u}_\mathrm{dust}$ are gas and dust velocities respectively.

As we show in section \S~\ref{sec:terminal_vel_and_limitation}, in the parameter regime of interest to us, the dust velocity is close to the ``terminal velocity'' (the velocity that enables the gas drag to balance all other forces acting on a grain), especially in the disk where most of the grain growth occurs. We are thus motivated to adopt the terminal velocity approximation, where the dust velocity components are given by
\begin{equation}
    \mathbfit{u}_\mathrm{dust} = \mathbfit{u}_\mathrm{gas} + \mathbf{\zeta} t_s
    \label{equ:vel_terminal}
\end{equation}
where the vector $\mathbf{\zeta}$ is the sum of all forces except the gas drag acting on a unit dust mass (i.e., the net dust  acceleration to be balanced by gas drag force per unit mass), with its three components given by
\begin{equation}
\begin{split}
    & \zeta_{r} = g_r + \frac{u_\mathrm{\phi, gas}^2 + u_\mathrm{\theta, gas}^2}{r} \\
    & \zeta_\theta = g_\theta - \frac{u_\mathrm{r, gas}u_\mathrm{\theta, gas} - u_\mathrm{\phi, gas}^2\cot\theta}{r} \\
    & \zeta_\phi = g_\phi - \frac{u_\mathrm{r, gas}u_\mathrm{\phi, gas} + u_\mathrm{\phi, gas}u_\mathrm{\theta, gas}\cot\theta}{r}
\end{split}
\label{equ:vel_terminal_zeta}
\end{equation}
This approximation is equivalent to the steady-state assumption adopted by \citet{Nakagawa1986} for the dust dynamics. It greatly simplifies the treatment of the grain dynamics, enabling us to focus on grain growth, which is more computationally expensive.

\subsubsection{Grain spatial transportation}
\label{sec:Grain_spatial_transportation}

The grain spatial transportation is modeled with the advection equation
\begin{equation}
    \frac{\partial N}{\partial t} = - \nabla\cdot(N \mathbfit{u})
    \label{equ: dn/dt original}
\end{equation}
where $N$ and $\mathbfit{u}$ are local dust number density and dust velocity respectively. We ignore diffusion because the thermal speed of all grains are less than a few cm/s, which is negligible comparing to their bulk speed ($>10^3$ cm/s). For the $i^{th}$ grain size bin in a cell located at $(I, J)$ ($\hat{r}$ direction indexed by $\{I\}$ and $\hat{\theta}$ indexed by $\{J\}$), we evolve the number density by solving the discretized advection equation at cell edges:
\begin{equation}
    \Delta N_{i}^{(I, J)} = \Big(\sum_{\mathrm{edges}} \mathrm{Flux}\Big)\Delta t
\end{equation}
In our models, there are two edges in the $\hat{r}$ direction and two edges in the $\hat{\theta}$ direction. The flux through the radial boundary between cells indexing $(I - 1, J)$ and $(I, J)$ is
\begin{equation}
    \mathrm{Flux}_{I-\frac{1}{2}} = 
    \Bigg\{
    \begin{array}{ll}
        u_{r, i}^{(I-\frac{1}{2},J)} A^{(I - \frac{1}{2}, J)}N_i^{(I-1, J)}, & \mathrm{if}\; u_{r, i}^{(I-\frac{1}{2},J)} > 0 \\
        u_{r, i}^{(I-\frac{1}{2},J)} A^{(I - \frac{1}{2}, J)}N_i^{(I, J)}, & \mathrm{if}\; u_{r, i}^{(I-\frac{1}{2},J)} < 0.
    \end{array}
\end{equation}
where $u_{r, i}^{(I-\frac{1}{2},J)}$ is the interpolated radial speed at cell edge and $ A^{(I - \frac{1}{2}, J)}$ is the area of the cell edge. 

Similarly the flux through the radial boundary between cells indexing $(I, J)$ and $(I+1, J)$ is
\begin{equation}
    \mathrm{Flux}_{I+\frac{1}{2}} = 
    \Bigg\{
    \begin{array}{ll}
        - u_{r, i}^{(I+\frac{1}{2},J)} A^{(I + \frac{1}{2}, J)}N_i^{(I, J)}, & \mathrm{if}\; u_{r, i}^{(I+\frac{1}{2},J)} > 0 \\
        - u_{r, i}^{(I+\frac{1}{2},J)} A^{(I + \frac{1}{2}, J)}N_i^{(I + 1, J)}, & \mathrm{if}\; u_{r, i}^{(I+\frac{1}{2},J)} < 0.
    \end{array}
    \label{equ:advectionr-to-r+1}
\end{equation}
where $u_{r, i}^{(I+\frac{1}{2},J)}$ is the interpolated radial speed at cell edge and $ A^{(I + \frac{1}{2}, J)}$ is the area of the cell edge. Note the leading negative sign in equation~\ref{equ:advectionr-to-r+1} reflects the causation that a positive flux from cell $(I, J)$ to cell $(I + 1, J)$ results in a net lost in cell $(I, J)$.
Flux in the $\hat{\theta}$ direction can be calculated with the same method. The algorithm is similar to the one used in \citet{Birnstiel2010} (equation~A4, A5 within).

At the boundaries in the radial direction, the speed at each boundary is taken to be the value at the nearest cell if the flow is outbound from the simulation domain and zero if the flow is inbound. At the boundaries in the $\theta$ direction, the speed at each boundary is taken to be 0 because of symmetry.

\subsubsection{Grain growth}
\label{sec:grain_growth_setup}

Grain growth is calculated within each cell at each time step with the Smoluchowski equation \citep{Smoluchowski1916} following \citet{Birnstiel2010}. We adopted the Smoluchowski equation for re-distributing grain species $i, j, k$ within a cell: 

\begin{equation}
    \frac{\partial n_k}{\partial t} = \iint_0^{\infty}M_{ijk}n_i n_j ds_ids_j
    \label{equ:s_eq}
\end{equation}
where the kernal $M_{ijk}$ is defined as
\begin{equation}
\begin{split}   
    M_{ijk} = &\frac{1}{2}K_{ij}\delta(m_i+m_j-m_k) - K_{ij}\delta(m_j-m_k) \\
    &+ \frac{1}{2}L_{ij}S_{ijk} - L_{ij}\delta(m_j-m_k)
\end{split}
\label{equ:coag}
\end{equation}
where $m_i, m_j$ and $m_k$ are the masses of individual grains and $K_{ij}$ and $L_{ij}$ are the coagulation kernel and fragmentation kernel respectively. $S_{ijk}$ describes the distribution of fragments in a collision. In reality, the best description of grain coagulation and fragmentation is to use a function describing the distribution of grain mass in the aftermath of a collision \citep[see e.g.][]{Hasegawa2021, Kobayashi2010}. As a first step to understand the grain growth in an actively forming protostellar disk, we have included coagulation but not fragmentation, which should yield an upper limit to the growth. We emphasize this idealized assumption by separating perfect sticking from all other processes in equ. \ref{equ:coag}. As we will see in \S~\ref{sec:dust_the_fiducial_model} below, even with this optimistic assumption, it is difficult to grow a substantial amount of large mm/cm grains in the simplest, baseline case. Including fragmentation would exacerbate the difficulty. 

Under the assumption above, we have 
\begin{equation}
    K_{ij} = \chi\cdot\Delta u_{ij}\sigma_{\mathrm{geo}, ij} = \chi\cdot \Delta u_{ij}\cdot\pi(s_i + s_j)^2
    \label{equ:growth_equation}
\end{equation}
\begin{equation}
    L_{ij} = 0
\end{equation}
where $\Delta u_{ij}$ is the relative speed between grains in bin $i$ and $j$; $\sigma_{\mathrm{geo}, ij}$ is the geometric cross-sectional area between grains of size $s_i$ and $s_j$. 
$\Delta u_{ij} = |\mathbfit{u}_i - \mathbfit{u}_j|$ describes the relative speed between grains in two size bins with speeds $\mathbfit{u}_i$ and $\mathbfit{u}_j$, whose values are taken to be grain speeds at the cell centers.

Note that we used only the grain drift velocity relative to the gas due to the body force in computing the relative speed for grain coagulation. This relative speed can potentially be greatly increased by effects not included in our calculations, such as turbulence \citep{Voelk1980, MMV1991, OrmelCuzzi2007}. To account for such possibilities, we have included an enhancement factor $\chi$ in the equation~(\ref{equ:growth_equation}) for the coagulation rate, with $\chi=1$ corresponding to the lower limit for the grain growth. We show in the Appendix that the Brownian motion does not affect the growth to large grains significantly.

In the numerical model, the Smoluchowski equation must be discretized as
\begin{equation}
    \frac{\Delta N_k}{\Delta t} = \sum_{ij}M_{ijk}N_i N_j
    \label{equ:s_eq}
\end{equation}
where $N_i$, $N_j$ and $N_k$ are the number density of grain species $i, j$ and $k$ respectively (equation~\ref{equ:defineN_ibyn_i}). Because of the discretization of grain masses, $m_i + m_j$ can fall between two consecutive bins with $m_{k}$ and $m_{k+1}$. In this case we modify the delta function in the kernel (equation~\ref{equ:coag}) to distribute the mass into both size bins in the following way: 
\begin{equation}
    \delta(m_i+m_j-m_k) \rightarrow \Bigg\{
    \begin{array}{ll}
        \frac{m_i + m_j - m_k}{m_{k+1} - m_k}, & \mathrm{if}\; m_k \leq m_i + m_j \leq m_{k+1} \\
        \frac{m_i + m_j - m_{k-1}}{m_k - m_{k-1}},  & \mathrm{if}\; m_{k-1} \leq m_i + m_j \leq m_{k} \\
        0, & \mathrm{otherwise}.
    \end{array}
\end{equation}

This modified $\delta$ function ensures mass conservation and suppresses the numerical ``leak'' of grains to larger size \citep[see e.g.][]{Lombart2021}. Because grain growth rate is significant only in and around the disk, where the density is highest, we perform grain coagulation calculations only in cells where hydro density is higher than $4\cdot10^{-17}\ \mathrm{g/cm^3}$. This region includes the entire disk at all times.

\subsection{Lagrangian Treatment of Dust Particles} 
\label{sec:dust_Lagrangian_method}

To verify the velocities obtained by the terminal velocity approximation (equation~\ref{equ:vel_terminal}) is a reasonable estimate of the grain speeds at the cell locations, we perform a series of simulations with the Lagrangian dust particles, and compare the speeds obtained with the terminal velocity approximation. Similar to the Eulerian dust simulation outlined in the last subsection, the Lagrangian particle simulations are done by post-processing the hydrodynamic simulation.

Each particle in the Lagrangian simulation is independently evolved in a 3D Cartesian coordinate. The particles move within the gas under the influence of both gravity and hydrodynamics drag force. The location of each particle is updated at every time step, with
\begin{equation}
    \Delta\mathbfit{x} = \mathbfit{u}_\mathrm{par}\Delta t + \frac{1}{2}\mathbfit{a}\Delta t^2
    \label{equ:lagrangian_evolve}
\end{equation}
where $\Delta t$ is the integration time step, $\mathbfit{u}_\mathrm{par}$ is the dust velocity at the last time step and $\mathbfit{a}$ is the instantaneous acceleration, given by
\begin{equation}
    \mathbfit{a} = \mathbfit{g} + \mathbfit{a}_\mathrm{drag} = \mathbfit{g} + \frac{\mathbfit{u}_\mathrm{gas} - \mathbfit{u}_\mathrm{par}}{t_s}
\end{equation}
where $\mathbfit{g}$ is the local gravitational acceleration, $\mathbfit{u}_\mathrm{gas}$ is the local gas velocity, $\mathbfit{u}_\mathrm{par}$ is the velocity of the particle, and $t_s$ is the stopping time. The stopping time can be very small for the smallest grains in the highest density regions of the disk, which severely limits the time step $\Delta t$ that is used to evolve all Lagrangian particles at the same time. To speed up the computation, we force those particles with stopping times smaller than a small floor value of $t_{s,\mathrm{min}}=0.01$~yrs to move together with the gas, which is physically reasonable since such particles are essentially tracer particles of the gas. This approximation does not affect the main purpose of the Lagrangian simulation, which is to test the validity of the terminal velocity approximation for relatively large particles (e.g., mm/cm sized grains; see \S~\ref{sec:terminal_vel_and_limitation} below), which tend to have stopping times much larger than 0.01~years.
%


The local hydrodynamical quantities ($\rho$, $T$, $\mathbfit{g}$, $\mathbfit{u}_\mathrm{gas}$) at the locations of the dust particles are obtained using the Triangular-Shaped-Cloud (TSC) algorithm \citep{TSC} and projected onto the Cartesian coordinate. 

The Lagrangian particle simulations are initialized at the same time as the Euler simulations. At initialization, we put 400,000 particles with the same size in the simulation domain between 10 and 8000 au with pseudo-random locations, with proportionally more particles at locations where the gas density is higher. The particles are then evolved spatially with integration method described by equation~(\ref{equ:lagrangian_evolve}).

\section{Hydro results}
\label{sec:hydro_results}

We start with a discussion of the gas dynamics of the core collapse and disk formation and evolution, which forms the basis for our grain growth calculations. 

\begin{figure}
    \centering
    \includegraphics[width=\columnwidth]{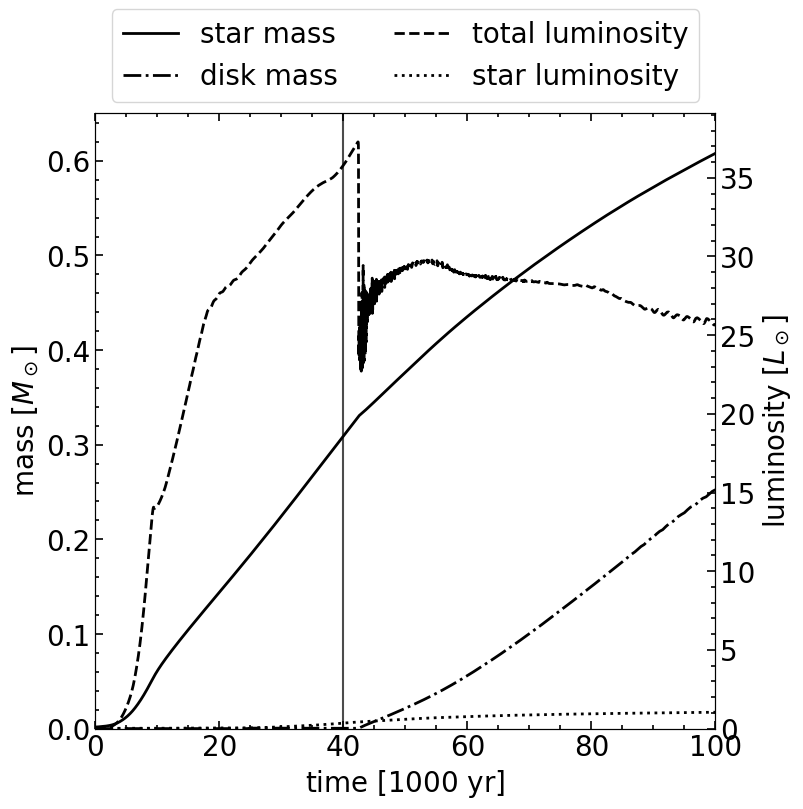}
    \caption{Stellar properties and disk mass during the gas simulation. The solid and dash-dotted lines are the star mass and disk mass respectively; dotted line and dashed line are the star luminosity ($L_*$) and total luminosity ($L$ in equation~\ref{equ:tot_luminosity}) respectively. The vertical line at $t=40,000$ years shows the starting time of the dust simulations.}
    \label{fig:hydro_star}
\end{figure}

Fig.~\ref{fig:hydro_star} shows the mass and luminosity of the central protostar in our model. The vertical line shows the stellar properties at the time ($t=40,000$~yrs) when the dust simulation was initialized. During the simulation of dust grains, the star grows from $\sim 0.3$ to $0.6~ M_\odot$. The total (stellar and accretion) luminosity stays between 25 and 30 $L_\odot$ after the disk forms; the sudden drop in luminosity at $\sim 42000$ years is caused by the reduction of mass accretion rate onto the central star due to disk formation. 

\begin{figure}
    \centering
    \includegraphics[width=\columnwidth]{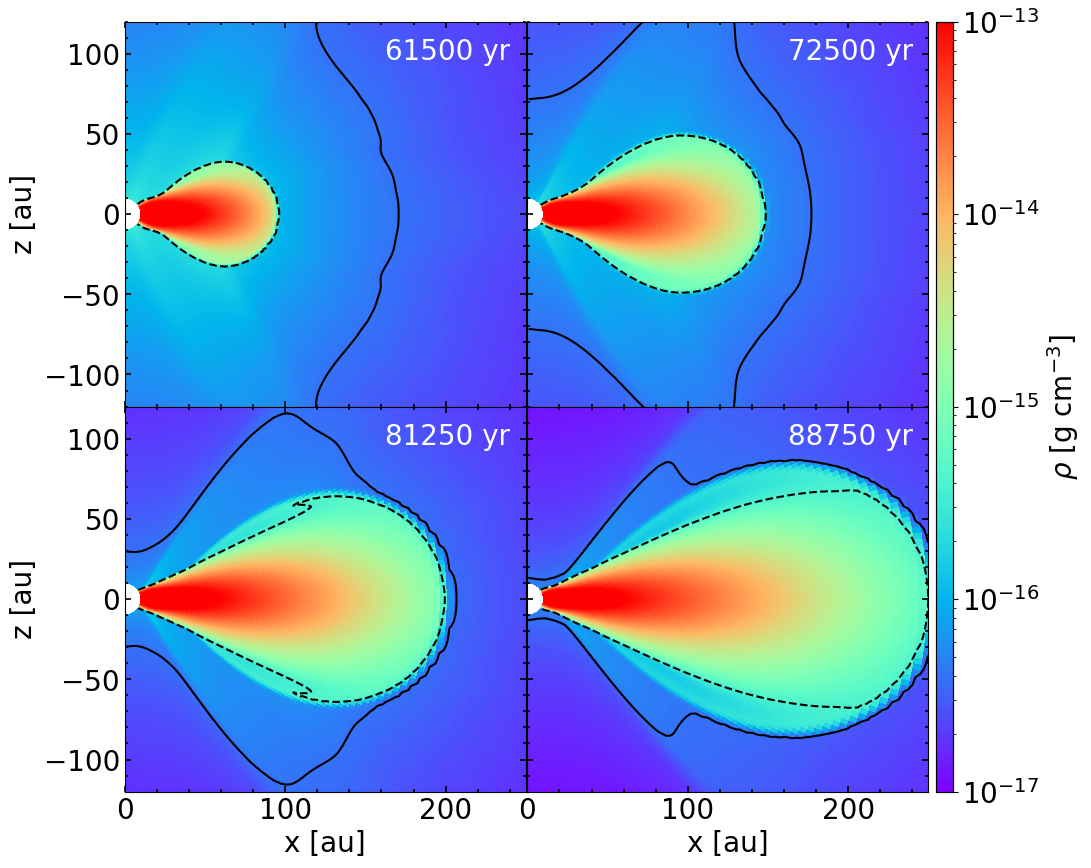}
    \caption{Hydrodynamical disk densities. Four panels are the disk density when the disk is approximately $100$, $150$, $200$ and $250$~au. Corresponding times of the frames are shown in each panel. The black dashed line is the location where gas density is $4\times 10^{-16}\mathrm{g\ cm^{-3}}$, where our definition of the boundary of the disk is located. The solid black line is the boundary in which dust coagulation is calculated, where gas density is $>4\times10^{-17}\mathrm{g\ cm^{-3}}$. 
    }
    \label{fig:hydro_disk_2D}
\end{figure}

Fig.~\ref{fig:hydro_star} also shows the mass of the ``disk'', defined to be anywhere with the gas density $>4\cdot10^{-16}\ \mathrm{g\ cm}^{-3}$. In fig.~\ref{fig:hydro_disk_2D} we show the meridian density distributions at four representative times, when the disk size is, respectively, 100, 150, 200 and 250~au, with the ``disk'' boundary marked by the black dashed contour in each panel. Our analysis of the dust below will focus on the latter three times. We choose the $4\cdot10^{-16}\ \mathrm{g\ cm^{-3}}$ cut-off density because its location is close to a sudden density increase from the rapidly infalling ``envelope'' to the ``disk'' at most times during the simulation. The solid black contour is the boundary in which dust coagulation is calculated, located at a gas density $4\cdot10^{-17} \ \mathrm{g\ cm^{-3}}$. The region is chosen to fully enclose where dust coagulation happens rapidly. Dust coagulation outside occurs at a negligible rate that can be ignored.

\begin{figure}
    \centering
    \includegraphics[width=\columnwidth]{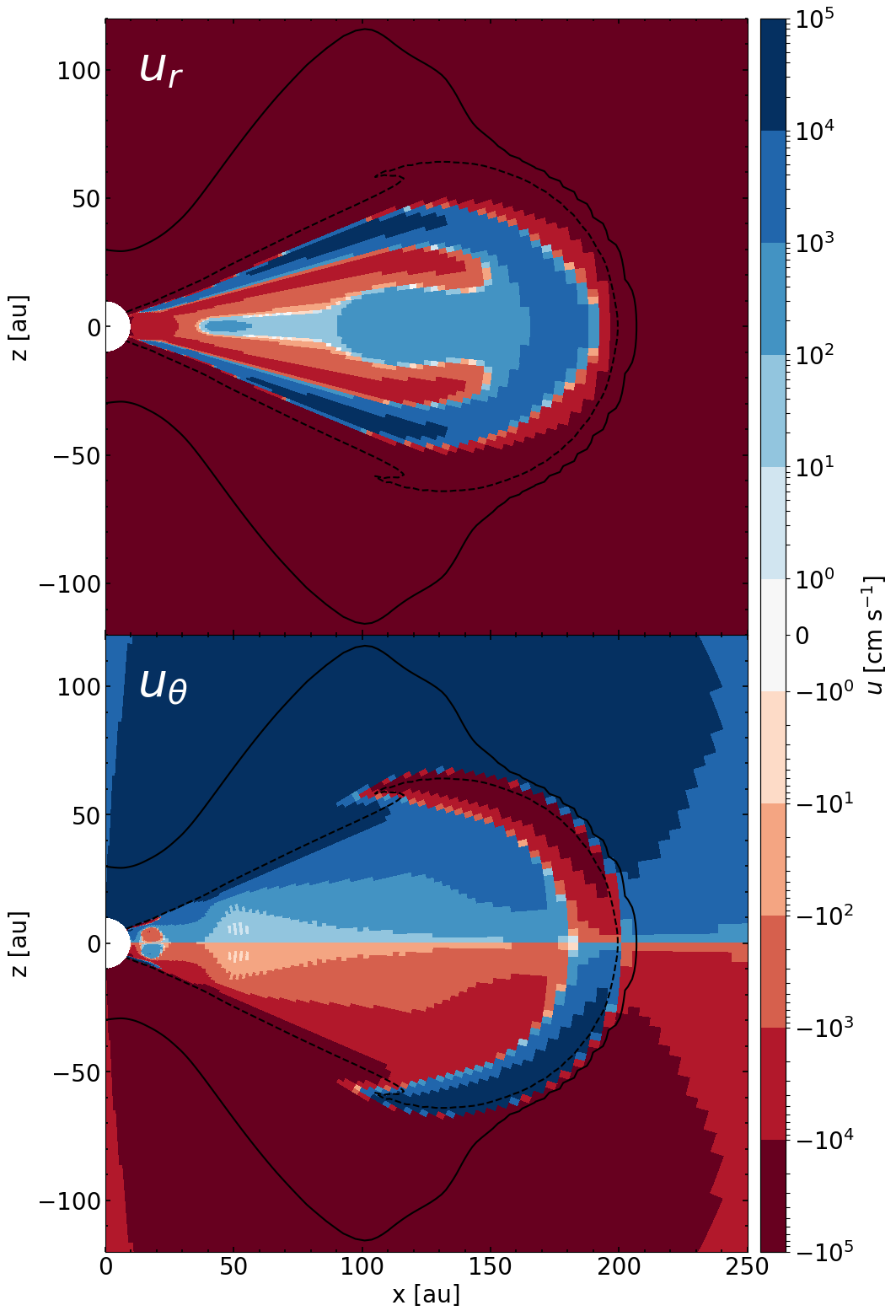}
    \caption{The speeds in $\hat{r}$ direction and $\hat{\theta}$ direction are shown in the upper and lower panels respectively. The figure is taken when the disk radius is about 200~au. At other times in the simulation when there is a disk, the velocity field is similar to the ones showing here. 
    }
    \label{fig:hydro_disk_2D_vr_vtheta}
\end{figure}

The meridional gas velocities in the $\hat{r}$ and $\hat{\theta}$ directions are shown in fig.~\ref{fig:hydro_disk_2D_vr_vtheta}. Gas velocities at other times are similar to the selected frame. We observe that $u_r$ is positive in some parts of the disk due to (viscous) angular momentum transport. In the majority of the disk, $u_\theta$ is pointing towards the midplane, with the exception at the innermost part of the disk and the edge of the disk. At the innermost part of the disk, an outgoing flow at the edge of the disk ``shears'' the in-going disk midplane, causing a circular motion at $\sim 20$ au. This feature can be numerical due to the existence of an inner boundary at $10$ au. At the outer (radial) edge of the disk, quickly infalling envelope encounters the expanding disk, forcing some of the gas to go around the disk and to move away from the midplane. 


\begin{figure}
    \centering
    \includegraphics[width=\columnwidth]{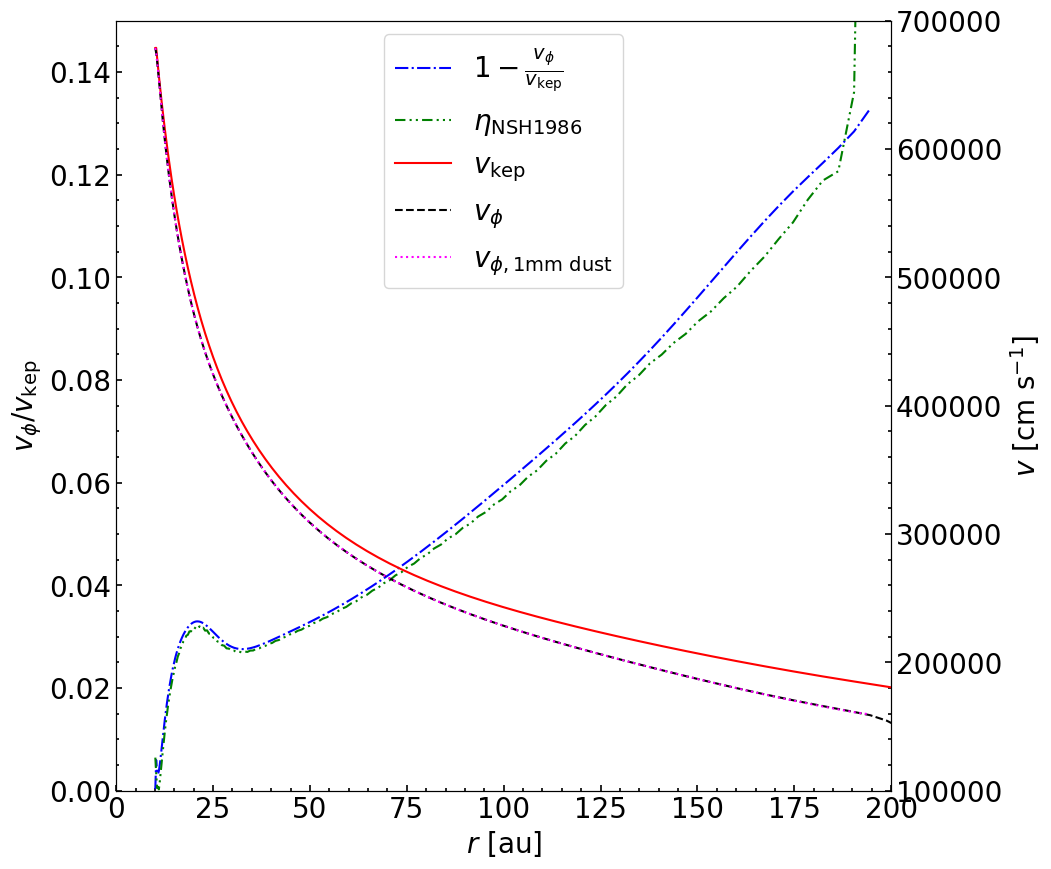}
    \caption{Sub-Keplerian rotation. Plotted are the rotation speed near the disk midplane, the local Keplerian speed (including the contribution from the self-gravity), and the fractional deviation of the azimuthal speed from the local Keplerian speed when the disk radius is about 200~au (solid line). The fractional deviation is close to the analytic estimate from \citet[][labeled ``NSH1986"]{Nakagawa1986}. 
    }
    \label{fig:vphi}
\end{figure}

As expected, the gas in the disk rotates at a speed below the local Keplerian speed because of the outward pressure gradient force. In Fig.~\ref{fig:vphi}, we plot the rotation speed and the local Keplerian speed as well as the fractional deviation from the local Keplerian speed on the midplane, $\vert u_\phi-v_K\vert/v_K$, at the time
when the disk radius is about 200~au. For comparison, we have plotted the analytic prediction from \citet{Nakagawa1986}. The slight difference between the numerical and analytical values are due to the inclusion of self-gravity in our simulation. As is well known, this sub-Keplerian gas rotation is the cause of the inward radial dust migration, which plays a key role in the grain growth and loss near the disk midplane.

\section{Dust Results}
\label{sec:dust_result}

Because dust coagulation is only modeled in the Eulerian models, we focus on the Eulerian results in this section. The Lagrangian results will be discussed in section \S~\ref{sec:terminal_vel_and_limitation} in connection with the terminal velocity approximation.

\subsection{The Baseline Model}
\label{sec:dust_the_fiducial_model}

\begin{figure}
    \centering
    \includegraphics[width=\columnwidth]{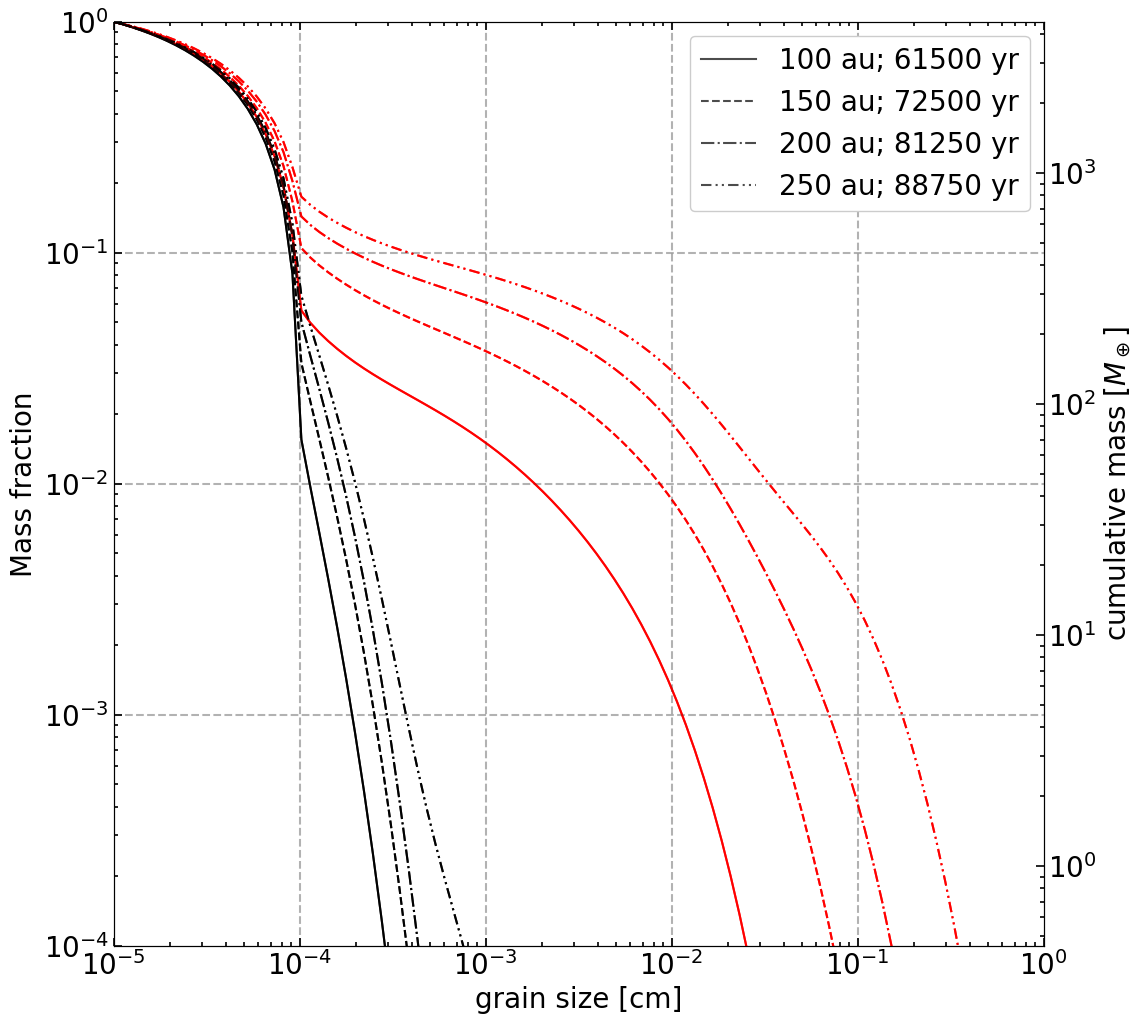}
    \caption{Cumulative grain mass as a function of the grain size. The mass includes both the grains in the active computational domain and those already advected through the inner boundary at $r = 10$~au. The total grain mass is conserved throughout the simulation, with a value of 4520 $M_\oplus$ determined at the beginning of the dust model (40,000 years). At later times, the total mass of the grains is calculated by summing all mass that has gone through the central boundary over time and adding all mass in the active computation domain (10 to 10,000~au).
    The value at each grain size is the mass fraction of the grains at or larger than that size. The black lines shows the cumulative grain mass at four different disk sizes and times in the baseline model ($\chi = 1$ in equation~\ref{equ:growth_equation}). The red lines are the cumulative values at the same times as the black lines with the same line style, but for the $\chi = 4$ model.
    }
    \label{fig:74_34-74_32_summary_plot}
\end{figure}

In the baseline model, we only consider the bulk drift motion of the dust grains in the laminar disk without any sub-grid model to enhance the grain-grain collision speed (i.e., setting the enhancement factor $\chi$ in equation~(\ref{equ:growth_equation}) to $\chi = 1$).   In fig.~\ref{fig:74_34-74_32_summary_plot} we show the cumulative mass and mass fraction of the grains from large to small, including all the mass in the simulation domain since the initialization of the dust simulation (that includes those going into the center during the simulation). Mass is well-conserved in the simulation with a total mass of 4520 $M_\oplus$. Fig.~\ref{fig:74_34-74_32_summary_plot} shows that about 8\% (by mass) of the dust has grown to sizes beyond the original maximum size of 1~$\mu$m by the time that the disk is 250~au in radius. However, at this time, less than 0.01\% of the dust has grown beyond 10~$\mu$m. This result demonstrates that it is difficult to produce a substantial population of large grains by relying  solely on the differential drift velocities of grains between different sizes to generate the relative grain-grain collision speed. The reason will be explored in detail in \S~\ref{sec:sol_to_smoluchowski} and \S~\ref{sec:discussion}. Here, we briefly state that the low grain growth rate is due to the initially micron-sized (or smaller) grains being well coupled to the relatively dense gas in the disk, which makes the grain-gas drift velocity (and thus the grain-grain collision speed) small and the grain growth slow.

\begin{figure*}
    \centering
    \includegraphics[width=\textwidth]{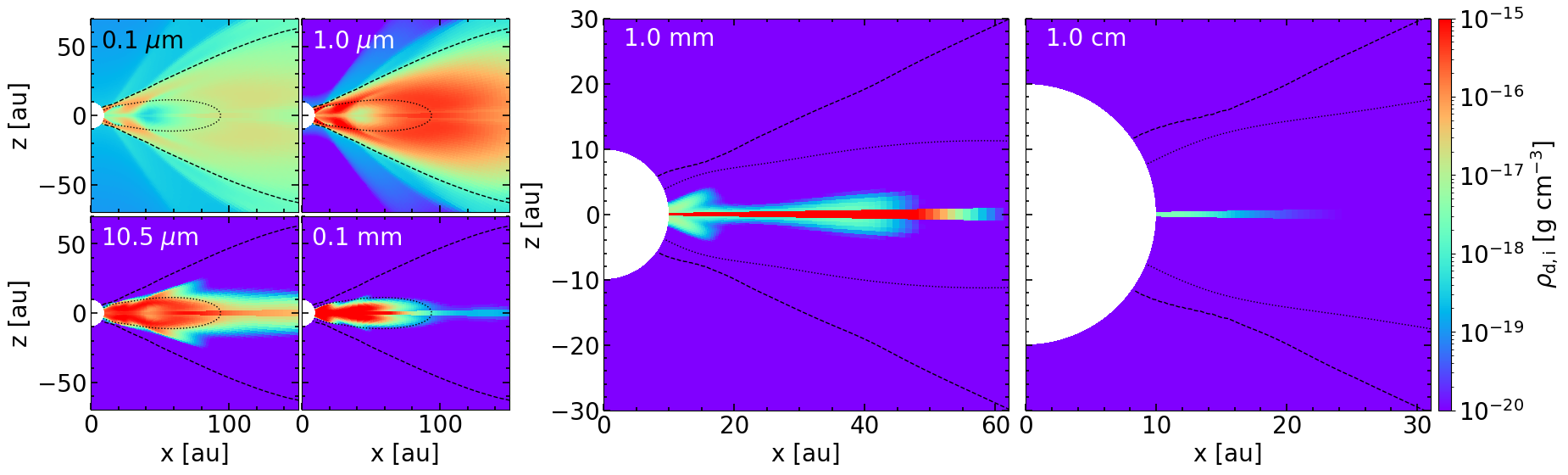}
    \caption{Spatial distributions of dust mass densities per logarithmic size decade of 6 size bins when the disk is at 250~au in the model with increased grain-grain collision speeds ($\chi = 4$ in equation~\ref{equ:growth_equation}). Each panel is the density of one size bin as noted in the upper left corner of each panel. Note the scale is different between the left four panels and the right two panels. We only show the inner disk regions in the two right panels since these are where the mm/cm-sized grains are concentrated. The black dashed contour is the boundary of the hydro disk, where the gas density is $4\times10^{-16}$g/cc. The dotted contour is at gas density of $4\times10^{-14}$g/cc to emphasize to scale difference between the left and right panels. Because we only simulated the upper hemisphere, the lower hemisphere is a reflection of the upper hemisphere about the midplane.} 
    \label{fig:74_10_dust_disk_rho}
\end{figure*}

\subsection{Model with Enhanced Grain-Grain Collision Speed}
\label{sec:dust_the_chi=4_model}

Since observational studies suggest that there might be a significant amount of large grains in protostellar disks in order to explain the dust emission at relatively long wavelengths (\S~\ref{sec:intro}), we are motivated to increase the grain growth rate over the baseline value $\chi=1$ in equation~(\ref{equ:growth_equation}). We experimented with different enhancement factors and found that $\chi = 4$ yields reasonable grain growth to large sizes.
Fig.~\ref{fig:74_34-74_32_summary_plot} shows the cumulative mass and mass percentage in the model with the increased growth rate. At the latest time shown (when the disk grows to 250~au), about 140~$M_\oplus$  of the initially small grains are converted to grains larger than 0.1~mm, and $\sim 15~M_\oplus$ has grown beyond 1~mm. 

Fig.~\ref{fig:74_10_dust_disk_rho} shows the spatial distributions of the mass densities per logarithmic size decade ($\rho_\mathrm{d, i}$) of dust grains of 6 representative sizes when the disk is at 250~au ($t=88750$~years). While the 0.1$\mu$m and 1$\mu$m sized grains are almost everywhere in the protostellar disk, grains larger than $10\mu$m are only concentrated at or around the disk midplane. The larger the grain size is, the smaller the vertical extent of its concentration becomes. Nearly all mm-sized grains are concentrated on the disk midplane, extending $\sim 50$~au in radius. The largest, cm-sized grains are concentrated within 20~au from the protostar on the disk midplane.

\begin{figure}
    \centering
    \includegraphics[width=\columnwidth]{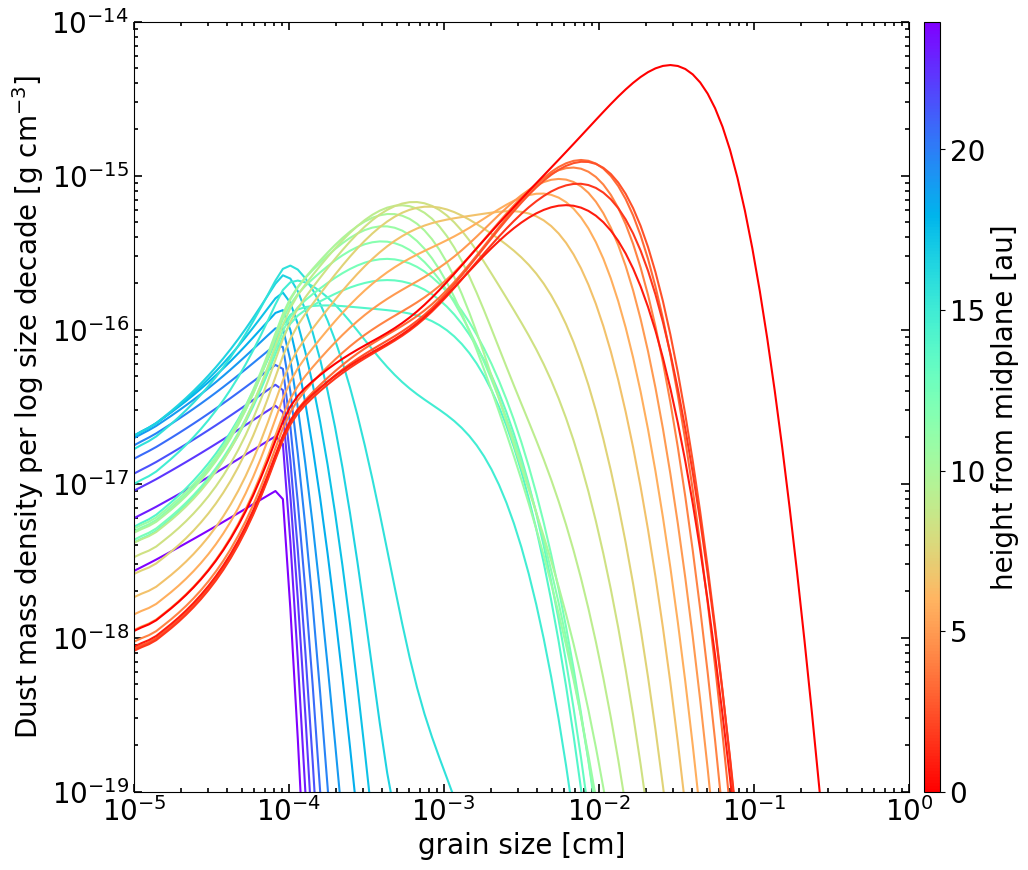}
    \caption{The distributions of the dust mass density per size decade in each logarithmic size bin as a function of the grain size at different heights from the midplane (one curve for each height, with the height specified by the colorbar) at a representative cylindrical radius of 50~au. This example frame is taken at $86250$ years, when the disk is at 250~au in radius. The highest red line, peaking around $0.5$~mm, shows that the largest grains are concentrated near the midplane.} 
    \label{fig:vertical_grain_distribution}
\end{figure}

To illustrate the substantial grain growth in this model more vividly, we plot in Fig.~\ref{fig:vertical_grain_distribution} the density per logarithmic size decade at different heights above the midplane at a representative cylindrical radius of $50$~au. 
Well above the midplane near the disk boundary, the size distribution is close to the initial MRN distribution. These are the ``raw material'' for later growth to larger grain sizes. As the grains move closer to the midplane, their distributions become more dominated by larger grains. Around a height $\sim 15$~au above the midplane, the size that dominates the grain mass (i.e., the peak of the distribution, $s_\mathrm{peak}$) suddenly increases from $\sim 1\ \mu$m ($10^{-4}$~cm) to $\sim 10\ \mu$m. Then $s_\mathrm{peak}$ grows more slowly with decreasing height. Closer to the midplane at a height between $\sim 1$ and 5~au, most of the grain mass resides in rather large grains of $\sim100\ \mu$m. Near the midplane, the distribution is dominated by the even larger, $\sim 0.5$~mm-sized grains (see the rightmost curve in Fig.~\ref{fig:vertical_grain_distribution}).  A similar increase in grain size towards the midplane is observed at other cylindrical radii, indicating wide-spread grain growth throughout the protostellar disk. In what follows, we seek to understand how the grain growth occurs in the simulation. 

\subsubsection{Interplay between drift and growth}
\label{sec:interplay_drift_growth}

Because both the gravity and the polar component of the gas velocity $u_\theta$ point towards the midplane, naturally the velocity vectors of most grains point towards the midplane in the majority of the disk. With the dust grains exerting no pressure force on each other, the midplane is a natural place for grains to concentrate. This concentration is the key to understanding the growth pattern that we see.

\begin{figure*}
    \centering
    \includegraphics[width=\textwidth]{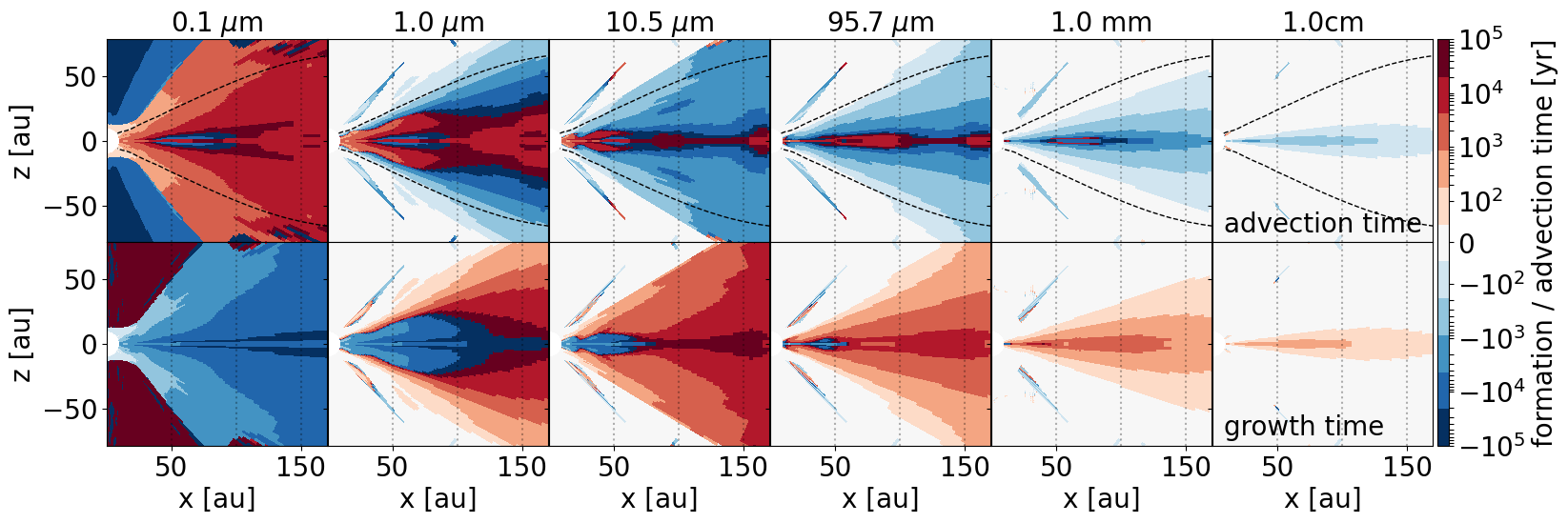}
    \caption{Grain advection and formation time scale for grains in six representative size bins when the disk reaches 250au. The upper six panels are advection time scale and the lower six panels are formation (growth) time scale. The defintion of the time scales is in section \S~\ref{sec:interplay_drift_growth}, equation~\ref{equ:t_adv} and \ref{equ:t_grow}. Small grains are being advected into the disk constantly, supplying the growth to larger grains in the protostellar disk. The gain due to formation time scale is shorter than the loss due to advection time scale for large grains in general.}
    \label{fig:advection_time_scale}
\end{figure*}

As shown in Fig.~\ref{fig:74_10_dust_disk_rho}, most of the largest grains are concentrated near the midplane. To understand whether such grains are formed in situ or advected from higher up in the disk, we first examine the advection and growth rates in the protostellar disk. In Fig.~\ref{fig:advection_time_scale} we show the advection time scale of the grains in six representative size bins at different times. The advection time scale for grains in a given size bin $i$ is defined as
\begin{equation}
    t_{\mathrm{adv}, i} = \frac{m_i}{f_i}
    \label{equ:t_adv}
\end{equation}
where $m_i$ is the total mass of the grains in the $i$th size bin within a cell and $f_i$ is the net flux of these grains through all boundaries of the cell. If $f_i > 0$, $t_\mathrm{adv,i}$ is the characteristic time scale for advection to fill the cell; if $f_i < 0$, $t_\mathrm{adv,i}$ is the characteristic time scale for advection to remove all grains within the cell. 

The formation (growth) time scale is defined in the same way, 
\begin{equation}
    t_{\mathrm{grow},i} = \frac{m_i}{(dm_i/dt)}
    \label{equ:t_grow}
\end{equation}
where $dm_i/dt$ is the net growth rate calculated from the Smoluchowski equation (equ.~\ref{equ:s_eq}). The time scales for other frames are similar to that of the frame shown.

For sub-micron sized grains ($<1\ \mu$m), the advection time scale on the disk is generally positive, which means that these small grains are effectively replenished by the infall from the envelope. These grains are the seeds for growing to larger grains. As the grain size increases, the vertical extent of the region with a positive advection time scale shrinks closer to the midplane (see top panels of  Fig.~\ref{fig:advection_time_scale}, from left to right). This is a reflection of the general tendency for the grains that have grown beyond the initial sizes to move towards the midplane, which leads to a net depletion of the grown grains at high altitudes (and hence a negative advection time scale) and a net gain of the grown grains at low altitudes (and hence a positive advection time scale). 

The transition from a net advective loss to a net advective gain moves closer to the midplane for larger grains, until the grains are large enough that their radial drift towards the central protostar prohibits this transition from happening. 
This effect is seen for the two largest grain sizes shown in Fig.~\ref{fig:advection_time_scale} (1~mm and 1~cm): advection leads to a net loss nearly everywhere in the disk, even at the disk midplane.

Quantitatively, for $10\ \mu$m sized grains and $100\ \mu$m sized grains, in the majority of the disk the advection time scale is between $-10^3$ and $-10^4$ years. This time scale is short compared to the simulation time scale of $4\cdot 10^4$ years, which means that the grains are effectively removed from the atmosphere of the disk and are being advected to the disk midplane or to the central protostar. The advection time scale for mm and cm-sized are shorter than those of the smaller size bins, meaning they are removed from the upper parts of the disk even more effectively.

To compare the formation time scale with the drift time scale, we point out that if $t_\mathrm{grow}$ and $t_\mathrm{adv}$ have the same sign (both negative or both positive), then the grains either are quickly lost from that location (both negative) or quickly accumulated (both positive). The more subtle cases are when their signs are different. If $t_\mathrm{grow} > 0 > t_\mathrm{adv}$ and $\vert t_\mathrm{adv} \vert > t_\mathrm{grow}$, then the presence of grains at that location is due to fast grain growth since the in-situ growth rate is faster than the advective removal rate. This is the case near the midplane for mm and cm-sized grains, and is generally true for 100~$\mu$m and 10~$\mu$m sized grains at relatively high altitudes as well. Qualitatively, this shows that the large grains, especially the largest ones, are formed in-situ more rapidly than they are removed. The ``raw materials'' (micron or sub-micron sized grains) that supply the growth process are progressively being advected from the envelope to the disk.

\begin{figure}
    \centering
    \includegraphics[width=\columnwidth]{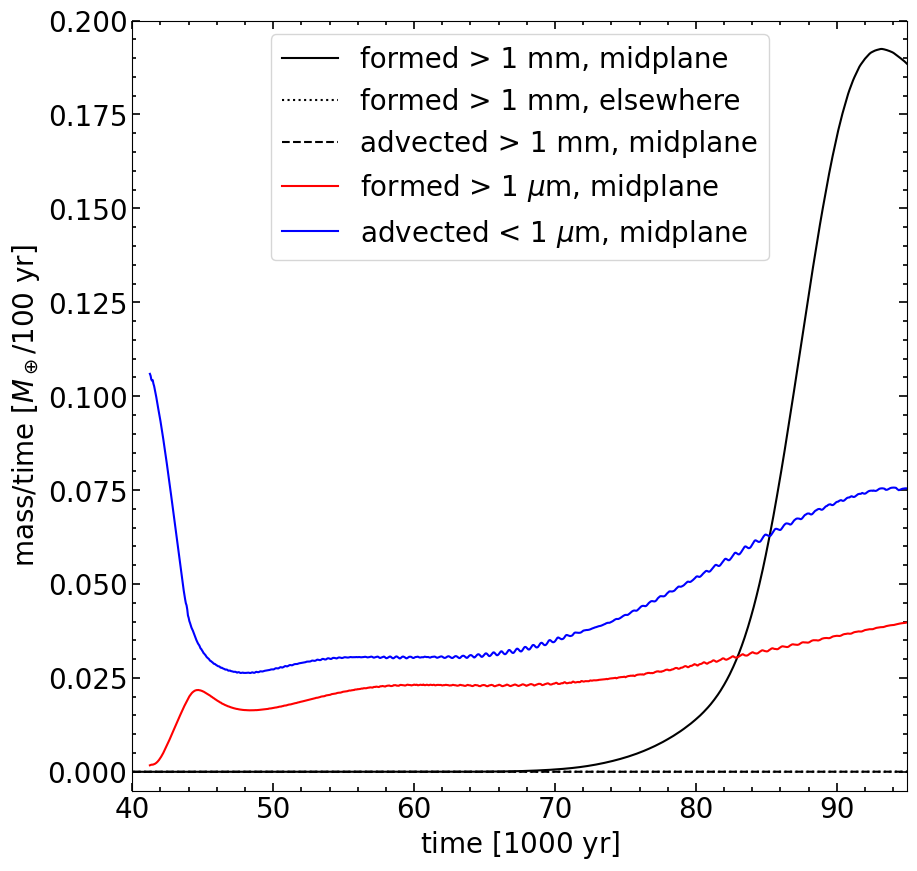}
    \caption{Comparison of in-situ formation rate and advection rate. The three black lines are for $>1$~mm-sized grains, showing that the in-situ formation rate in the midplane region (solid black line) is much larger than the rate with which such grains are advected into the midplane region (dashed black line) and the formation rate outside the midplane region (dotted black line; not visible due to overlap with the dashed line). Also plotted are the in-situ formation rate for $>1\ \mu$m-sized grains in the midplane region (red solid line) and the advection rate of $<1\ \mu$m grains into the midplane region (blue solid line), showing that more than half of the small ($<1\ \mu$m) grains from the initial MRN distribution advected into the midplane region are turned into larger ($>1\ \mu$m) grains.
    } 
    \label{fig:flux_to_midplane}
\end{figure}

To quantitatively examine whether the concentration of the largest grains near the midplane we observe in the density plots of Fig.~\ref{fig:74_10_dust_disk_rho} is due to the grains formed locally, we compare the total mass of $>$1~mm size grain advected to the midplane and the total mass of those that grew there locally. Fig.~\ref{fig:flux_to_midplane} shows the midplane formation rate of the grains larger than 1~mm (dashed line) and their formation rate anywhere else (dotted line), as well as the rate at which such grains are advected to the midplane (solid black line). Because the midplane surface in our simulation is a cell boundary instead of a cell center, we define the disk midplane to be one cell above or below the midplane surface and has a gas density greater than $4\times10^{-16}\mathrm{g\ cm}^{-3} $. The flux into this region is defined to be the total flux into the region but excluding the loss through the inner boundary at $r = 10$~au. We find that the amount of $>$mm size dust advected to the midplane and formed everywhere else are both negligible compared to the ones formed on the midplane locally. These results show that the largest grains that we see in the simulation are formed locally on the midplane instead of being advected from higher up in the disk.

In Fig.~\ref{fig:flux_to_midplane} we also plot the advection rate of all grains smaller than 1~$\mu$m and the formation rate of all grains larger than 1~$\mu$m on the midplane. Since mass is conserved during advection and formation, the advected small grains to the midplane are turned into the larger ones. The rate of growth into larger grains (>$1\ \mu$m) is about $60\%$ of the advection rate of the small grains almost at all times. In other words, more than half of the small grains are constantly being turned into larger ones. It supports a scenario where the small grains are carried to the midplane, converted to larger grains, which stay near the midplane or are advected radially to the central protostar.

\subsubsection{Dust-to-gas ratio}
\label{sec:g2d_dust_size_layer}

For the eventual formation of planets, one important quality to characterize is the dust-to-gas ratio since a high ratio is required to form planetesimals through streaming instability. We seek to understand the dust-to-gas ratio in our simulation. 

\begin{figure*}
    \centering
    \includegraphics[width=\textwidth]{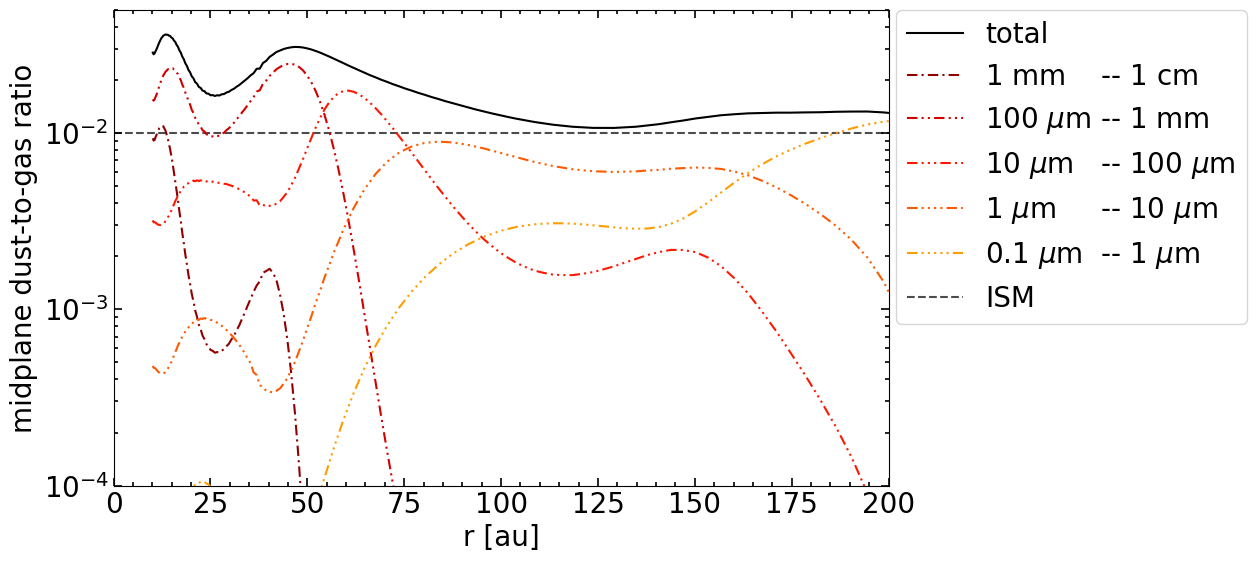}
    \caption{Dust-to-gas ratio on the disk midplane including grains with different sizes. The solid line is the dust-to-gas ratio including all dust grains. At around $\sim50$~au the dust-to-gas ratio reaches 0.03, three times the initial ISM value of 0.01 (dashed horizontal line). The colored lines are the dust-to-gas ratios on the midplane including only grains within a certain size decade. These colored lines show the contribution of dust in each size decade to the local dust-to-gas ratio. For example, $0.1\sim1$mm-sized grains contribute most to the dust mass around 45~au.
    }
    \label{fig:dust_to_gas_ratio}
\end{figure*}

In Fig.~\ref{fig:dust_to_gas_ratio} we plot the midplane dust-to-gas ratio as a function of radius at 85,000 years. The solid black line includes grains of all sizes. Each of the colored lines is the contribution of the grains in one size decade to the local dust-to-gas ratio. The total dust-to-gas ratio reaches $\sim0.03$ at a radius of $\sim50$~au, and the value decreases to slightly above the ISM value of 0.01 at and beyond $\sim125$~au. This increased concentration of dust grains will be important for our understanding of why grains grow on the disk midplane in \S~\ref{sec:sol_to_smoluchowski} and \S~\ref{sec:discussion}. 

In terms of the contribution by grains within each size decade, there is a ``layering'' of the highest contribution as a function of grain size and radius. Ignoring the peaks within 20~au which may be affected by the inner boundary, we find that the greatest contributor to dust-to-gas ratio near 45~au are the grains between $100\ \mu$m and 1~mm; the greatest contributor changes to grains between $10\ \mu$m and $100\ \mu$m at about 60~au. At 90~au the greatest contributor shifts to grains between $1\ \mu$m and $10\ \mu$m. Outside 160~au, the initial MRN grains dominate the contribution to the dust-to-gas ratio. With increased grain sizes, the region where the grains concentrate moves inward. The reason for this layering is the differential drift, with larger grains drifting inwards faster than smaller ones in the low density regions. Once the large grains reach a high density region, there is a ``traffic jam'' that concentrates them at those locations. This differential drift also explains why the largest (mm to cm-sized) grains do not contribute much to the dust-to-gas ratio at all radius: their radial drift is so fast that they hardly have time to accumulate in the disk. This quick radial drift can also be seen in Fig.~\ref{fig:advection_time_scale}, where the cm-sized grains are advected away on a time scale of $\sim1000$ years. If there are substructures, such as dense rings, in the gas disk, the large grains may be trapped on the disk. The dust trapping may make it possible to increase the local dust-to-gas ratio further, potentially to values that are conducive to the onset of the streaming instability \citep[$\sim0.1$, ][]{YoudinGoodman2005}. 

\section{Grain Growth Timescale}
\label{sec:sol_to_smoluchowski}

To understand the dust simulation results, we seek to obtain an estimate of the grain growth timescale using the Smoluchowski equation (equation~\ref{equ:s_eq}). In our case, the kernel $M_{ijk}$ is determined by the differential drift speed between grains of different sizes only. Using the terminal velocity approximation (equation~\ref{equ:vel_terminal}), we can write
\begin{equation}
    \Delta u_{ij} = \chi\zeta|t_{si} - t_{sj}| = \chi\zeta\frac{\tilde{\rho}_\mathrm{dm}}{\rho_\mathrm{g}v_\mathrm{th}}|s_i - s_j|
    \label{equ:delta_uij}
\end{equation}
where $\Delta u_{ij}$ is the relative speed defined in equation~(\ref{equ:growth_equation}), $\zeta = |\mathbf{\zeta}|$ is the magnitude of the net grain acceleration from all forces other than the gas drag that is balanced by the acceleration from the gas drag (equ.~\ref{equ:vel_terminal_zeta}); it depends only on the gas properties.  The quantity $t_{si}$ is the stopping time for the $i$-th dust size bin as defined in equation~(\ref{equ:stopping_time}). We include $\chi$ here for later discussions. Considering coagulation only, the net change of the number density of grains of a certain size $s_k$ is
\begin{equation}
    \begin{split}
        \frac{dn_k}{dt} &= \frac{1}{2}\iint_0^\infty \chi\zeta\frac{\tilde{\rho}_\mathrm{dm}}{\rho_\mathrm{g}v_\mathrm{th}}|s_i - s_j|(s_i+s_j)^2\delta_{ijk}n_in_jds_ids_j \\
        &\quad - \iint_0^\infty \chi\zeta\frac{\tilde{\rho}_\mathrm{dm}}{\rho_\mathrm{g}v_\mathrm{th}}|s_i - s_j|(s_i+s_j)^2n_in_j\delta_{jk}ds_ids_j
    \end{split}
    \label{equ:dndt_analytic}
\end{equation}
where the first term on the right hand side (RHS) describes the gain via coagulation and the second term describes the lost due to coagulation. $\delta_{ijk} = 1$ if $s_i^3 + s_j^3 = s_k^3$ and 0 otherwise; $\delta_{jk} = 1$ if $s_j=s_k$ and 0 otherwise. 
Assume the grain number density follows a power law $n_i(s_i) = As_i^{-q}$ from $s_s$ to $s_l$ (note that $q = 3.5$ is the MRN distribution). If the total grain mass density is given by $\rho_\mathrm{d, tot}$, the coefficient $A$ can be written as
\begin{equation}
    A = \frac{3(4-q)\rho_\mathrm{d, tot}}{4\pi\tilde{\rho}_\mathrm{dm}(s_l^{4-q} - s_s^{4-q})}
\end{equation}
A growth time-scale can be obtained by evaluating $n_k/(dn_k/dt)$. Each of the two $\delta$ functions in equation~\ref{equ:dndt_analytic} can be absorbed with one of the two integrals respectively, and we have
\begin{equation}
    \begin{split}
        \frac{dn_k}{dt} &= \frac{1}{2}\int_{s_s}^{s_k} \chi\zeta\frac{\tilde{\rho}_\mathrm{dm}}{\rho_\mathrm{g}v_\mathrm{th}}|s_j^\mathrm{gain} - s_i|(s_j^\mathrm{gain}+s_i)^2n_i n_j^{\mathrm{gain}} ds_i \\
        &\quad - \int_{s_s}^{s_l} \chi\zeta\frac{\tilde{\rho}_\mathrm{dm}}{\rho_\mathrm{g}v_\mathrm{th}}|s_j^{\mathrm{lost}} - s_i|(s_j^{\mathrm{lost}}+s_i)^2n_in_j^{\mathrm{lost}}ds_i
    \end{split}
    \label{equ:Nk_integration}
\end{equation}
where
\begin{equation}
    \begin{split}
        & s_j^{\mathrm{gain}} = [s_k^3 - s_i^3]^{1/3} \\
        & s_j^{\mathrm{lost}} = s_k \\
        & n_j^{\mathrm{gain/lost}} = A\Big(s_j^{\mathrm{gain/lost}}\Big)^{-q}
    \end{split}
\end{equation} 
Equation~\ref{equ:Nk_integration} can be evaluated by plugging in $n_i$, $A$ and simplify. Before writing out the result, we define one important parameter that governs the grain growth rate:
\begin{equation}
    \eta = \frac{4\rho_\mathrm{g}v_\mathrm{th}}{3\chi \rho_\mathrm{d, tot}\zeta}
    \label{equ:smoluchowski_coag_only_define_eta}
\end{equation}
and two expressions to simplify notations:
\begin{equation}
    w_\mathrm{gain/lost}(s_i, q) \equiv |s_j^\mathrm{gain/lost} - s_i|(s_j^\mathrm{gain/lost}+s_i)^2s_i^{-q}\Big(s_j^\mathrm{gain/lost}\Big)^{-q}
    \label{equ:F(s_i, q)}
\end{equation}
and
\begin{equation}
    f(s_k; s_s, s_l, q) \equiv \frac{s_k^{-q}({s_l^{4-q} - s_s^{4-q}})/({4 - q})}{\frac{1}{2}\int_\mathrm{gain} w_\mathrm{gain}(s_i, q)ds_i - \int_\mathrm{lost} w_\mathrm{lost}(s_i, q)ds_i}
    \label{equ:f(s_k; s_s, s_l, q)}
\end{equation}
where $\int_\mathrm{gain}$ and $\int_\mathrm{lost}$ are the integration bounds of the first and second term in equation~\ref{equ:Nk_integration} respectively.

Using these definitions, equation~\ref{equ:Nk_integration} can be written as a time scale
\begin{equation}
    \begin{split}
        t_\mathrm{depl} \equiv -\frac{dt}{d\ln n_k} = - \frac{n_k}{\frac{dn_k}{dt}} = - \eta\ f(s_k; s_s, s_l, q),
    \end{split}
    \label{equ:N/dNdt}
\end{equation}
Physically, $t_\mathrm{depl}$ is the time scale to deplete small grains of a given size ($k$) due to growth to larger sizes. It is positive for $\frac{dn_k}{dt} < 0$. As we show below, it is related to the time scale $t_\mathrm{size}$ to increase the grain size where the dust mass distribution peaks 
(see equation~\ref{equ:dlns/dt} below). 

Since $f(s_k; s_s, s_l, q)$ is a dimensionless quantity, the quantity $\eta$ has the unit of time, which can easily be confirmed from its definition in equation~(\ref{equ:smoluchowski_coag_only_define_eta}). It will be referred to as ``the characteristic timescale" hereafter. 
Note that all physical properties of the disk, such as its gas density $\rho_{\rm g}$, thermal speed ($v_\mathrm{th}$ (and thus temperature), grain acceleration from gas drag $\zeta$, as well as the dust-gas drift speed enhancement factor $\chi$, affect the grain growth only through the timescale $\eta$. If $\eta$ is increased (or decreased), the timescale for the grains to evolve would increase (or decrease) proportionally. 


In the next subsection (\S~\ref{sec:one-zone}) we will use a one-zone model to show numerically that there is a simple relationship between the timescale for the grain size to increase by a factor of $e$ (the size $e$-folding time $t_{\rm size}$) and the characteristic timescale $\eta$. We will then validate this relationship semi-analytically using equation~(\ref{equ:N/dNdt}) in the following sub-section (\S~\ref{sec:analytic_recipe}).
%

\subsection{One-Zone Model}
\label{sec:one-zone}

To quantify the role of the characteristic timescale $\eta$ in grain growth, we develop a so-called ``one-zone model," where we keep the hydro properties in a representative cell in the hydro simulation fixed and evolve the grain size distribution in time through coagulation from an initial distribution, keeping the dust-to-gas ratio to 0.01 (i.e.,  no outflow from or inflow into the cell). The algorithm for grain growth is identical to the one described in section \S~\ref{sec:grain_growth_setup}, but with two slight modifications: the number of grain bins is increased to 1050, 10 times the resolution used in the full model. This is to minimize possible numerical effects in our model due to the discretization of grain sizes \citep[see e.g.][]{Lombart2021}. The integration time step is chosen to be the smaller value between 1 year and the depletion time of a grain species. This adaptive time step is implemented to speed up the calculation with little loss of precision. For the rest of the discussion, the hydro background is chosen to be the disk at 85,000 years, when the disk size is between 200 and 250~au. 

\begin{figure}
    \centering
    \includegraphics[width=\columnwidth]{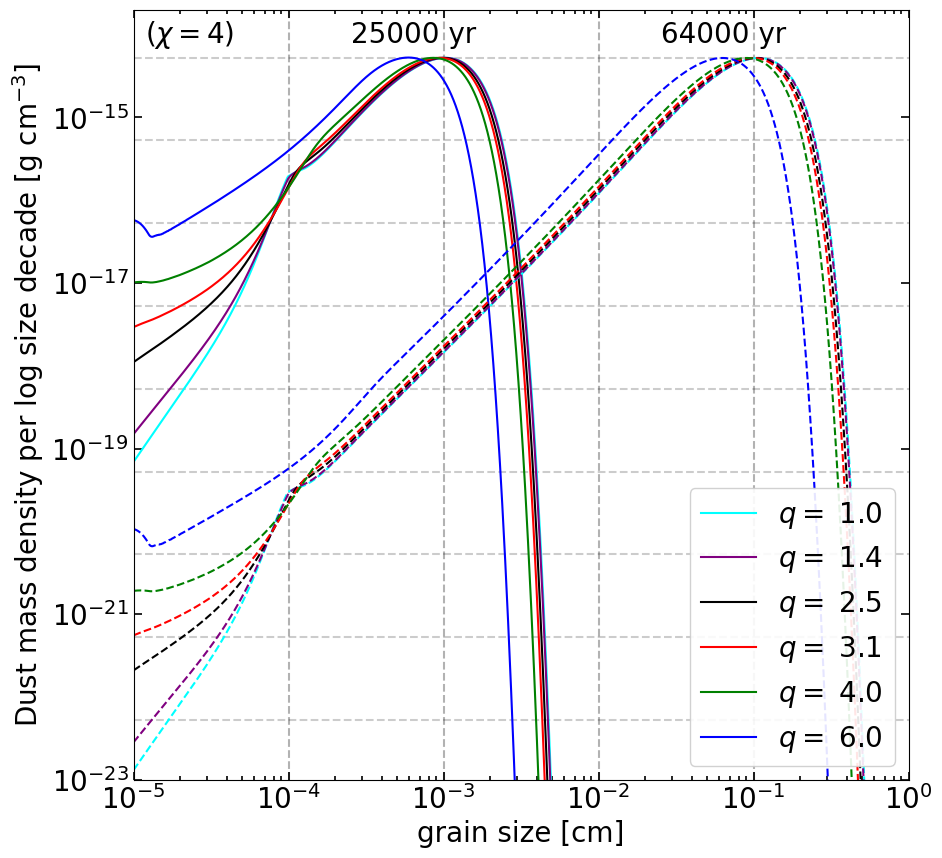}
    \caption{The distribution of grain mass density $\rho_{\rm d}$ per log size decade as a function of grain size at two times distinguished by line style: the solid line is taken at 25,000 years and the dashed line at 64,000 years. Each color denotes a different initial grain size power-law index $q$. The similarity of the curves at different times and for different initial size distributions indicates that the grain evolution is nearly self-similar in time and insensitive to the initial size distribution. The time for the grains to grow by one order in this example is about 19500 years (with $\chi = 4$). }
    \label{fig:one_cell_q_and_chi}
\end{figure}

We first study the effects of the initial grain size power-law index $q$ in the Smoluchowski equation (equation~\ref{equ:s_eq}, \ref{equ:coag} and \ref{equ:growth_equation}). We first take the hydro background to be the one cell on the midplane at 26~au (referred to as the ``fiducial cell'' hereafter). We integrate the Smoluchowski equation with different values of initial $q$ and plot the resultant dust mass density per size decade in Fig.~\ref{fig:one_cell_q_and_chi}. To speed up the calculations, we uses $\chi=4$ to conduct the one-zone models. The value chosen is identical to the $\chi$ used in the increased grain growth model (\S~\ref{sec:dust_the_chi=4_model}). 

The solid lines in Fig.~\ref{fig:one_cell_q_and_chi} are the solutions at 25,000 years after the initiation of grain growth and the dashed lines are at 64,000 years. Comparing models with different values of initial $q$, we find that the solutions are self-similar in time and share almost exactly the same shape (except in the $0.1\ \mu$m to 1~$\mu$m part, where all grains are initialized) regardless of the initial distribution. Although in the $q = 6$ case the distribution is lagging behind the other lines in moving to the larger size end by about 4000 years, the rate at which it is moving is the same as the other lines. These findings show that the solution to the Smoluchowski equation in a closed environment is a self-similar, attractive solution moving from small grain sizes to large grain sizes as a function of time. Therefore, we define the ``grain growth rate'' to be the rate that this self-similar and attractive solution moves to the larger grain side. 

In equation~\ref{equ:N/dNdt}, the part of the expression that is independent of the grain size $s$ and the size power-law index $q$ is the characteristic timescale $\eta$, as defined in equation~\ref{equ:smoluchowski_coag_only_define_eta}. The existence of a self-similar and attractive solution means that the grain growth rate is solely governed by the parameter $\eta$. For the purpose of later discussions, we approximate the dust mass density per logarithmic size decade $\rho_{d, i}$ as a function of $s_i$ as a power-law with index $p_s$: $\rho_\mathrm{d, i}\propto s_i^{p_s}$. From the one-zone models, we obtain $p_s\sim 1.9$. 

\begin{figure*}
    \centering
    \includegraphics[width=\textwidth]{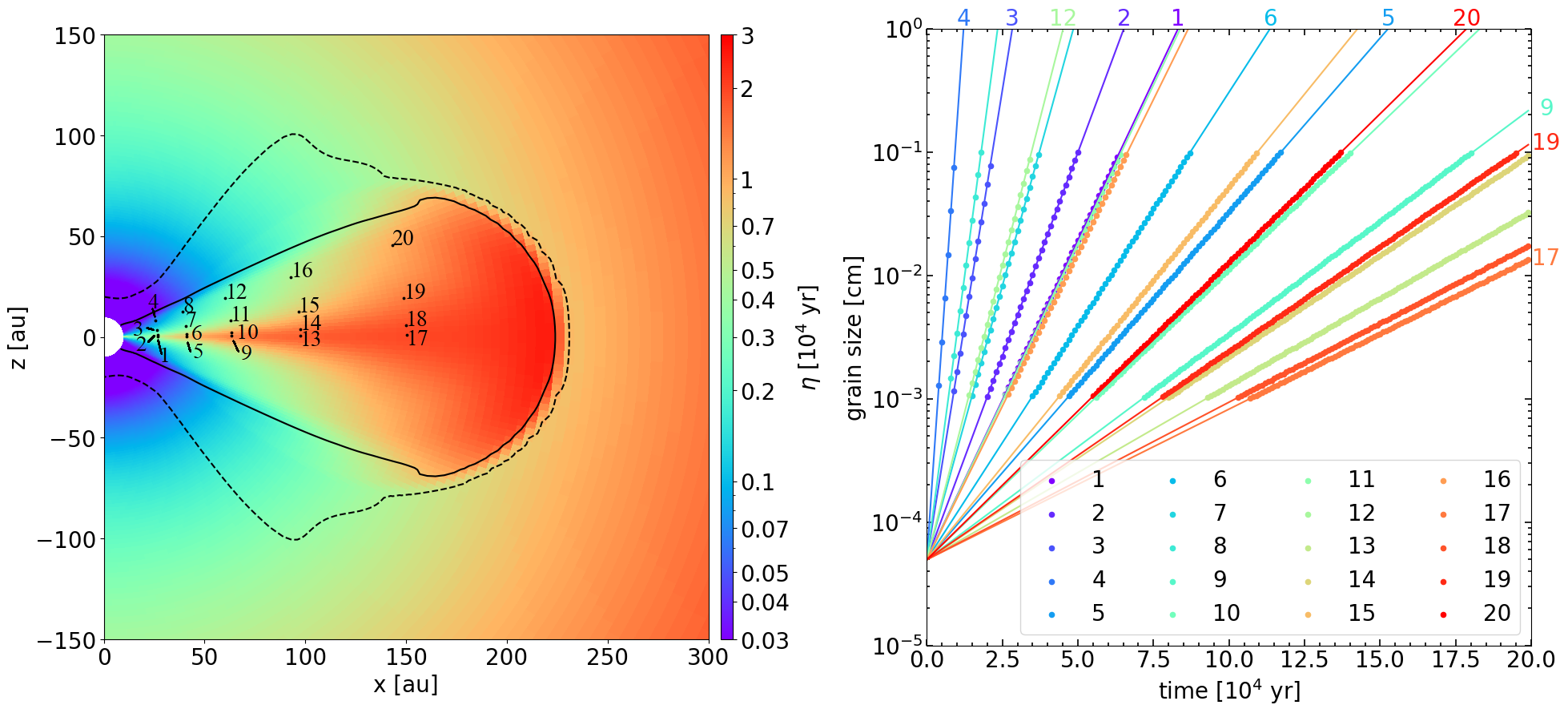}
    \caption{{\bf Left panel:} the distribution of the characteristic timescale $\eta$ in units of $10^4$~years assuming a dust-to-gas ratio of 0.01 everywhere and $\chi = 4$ and using the gas density $\rho_g$, thermal speed $v_\mathrm{th}$, and the acceleration from aerodynamic drag $\zeta$ at a representative time $t=85000$~years. {\bf Right panel:} Fitting of the peak grain size $s_{\rm peak}$ as a function of time using equation~\ref{equ:smax=10^(at+b)}. Each line corresponds to one location labeled in the left panel. The fitting agrees well the results of the one-zone simulations, proving the validity of equation~\ref{equ:smax=10^(at+b)}.
    }
    \label{fig:2D_eta}
\end{figure*}

Next, we seek to find a relationship between the characteristic timescale $\eta$, defined in equation~\ref{equ:smoluchowski_coag_only_define_eta} and the rate at which the self-similar distribution moves to the large grain side (the grain size increase rate). To find this relationship we pick 20 cells, distributed inside the disk at radii $26, 41, 63, 97$ and $150$~au and angles $0, 0.01\pi, 0.04\pi$ and $0.1\pi$ from the midplane. The cells are labeled in the left panel of Fig.~\ref{fig:2D_eta}, where the spatial distribution of the characteristic timescale $\eta$ (with the dust-to-gas ratio fixed at 0.01 and $\chi=4$) is plotted. The grain size distributions in the other 19 chosen cells are similar to that of the fiducial cell (located on the midplane at 26~au) shown in Fig.~\ref{fig:one_cell_q_and_chi}, but the grain growth rates are different. 

At each time, there is a grain size bin that has more mass than any other bins. We term the grain size in this bin ``the peak size" and denote it by $s_\mathrm{peak}$. For example, $s_\mathrm{peak}\sim 10~\mu$m in the fiducial cell at 25,000 years (see Fig.~\ref{fig:one_cell_q_and_chi}). In principle there are other ways to define a representative grain size, such as the mass weighted average size \citep{Kobayashi2016}. However, because the grain mass distribution is self-similar over time, our conclusion regarding the grain growth time scale is expected to be insensitive to how the representative grain size is chosen, as we verified explicitly for the case of mass weighted average size. We can fit the peak size as a function of time. Because at the beginning of the simulation the peak size is determined by the initial condition rather than grain growth, 
we only perform the fitting using integration results when the peak size is between $10~\mu$m and $1$~mm. We fit the peak size as a function of time to
\begin{equation}
    s_\mathrm{peak} = \exp\left\{\frac{t}{t_{\rm size}} + c\right\}
    \label{equ:smax=10^(at+b)}
\end{equation}
where $t_{\rm size}$ is the peak size growth timescale, which is to be fitted together with the constant $c$, and $t$ is the time. The fitting yields a pair of $t_{\rm size}$ and $c$ in each cell. Because the Smoluchowski equation is governed by the characteristic timescale $\eta$, we can fit $t_{\rm size}$ and $c$ as functions of $\eta$. We use a simple linear regression for the fitting of both. The best fit results are 
\begin{equation}
    t_{\rm size} = 1.84\cdot\eta
    \label{equ:smax_fit_a}
\end{equation}
and
\begin{equation}
    c = - 9.91.          
\end{equation}
Fig.~\ref{fig:2D_eta} shows that the peak size obtained from the fitted expression (equation~\ref{equ:smax=10^(at+b)}) agrees well with the numerical results from the one-zone simulations. In what follows, we will validate the relationship between $t_{\rm size}$ and $\eta$ (equation~\ref{equ:smax_fit_a}) semi-analytically.

\subsection{Semi-analytic Estimate of Grain Growth time}
\label{sec:analytic_recipe}

\begin{figure}
    \centering
    \includegraphics[width=\columnwidth]{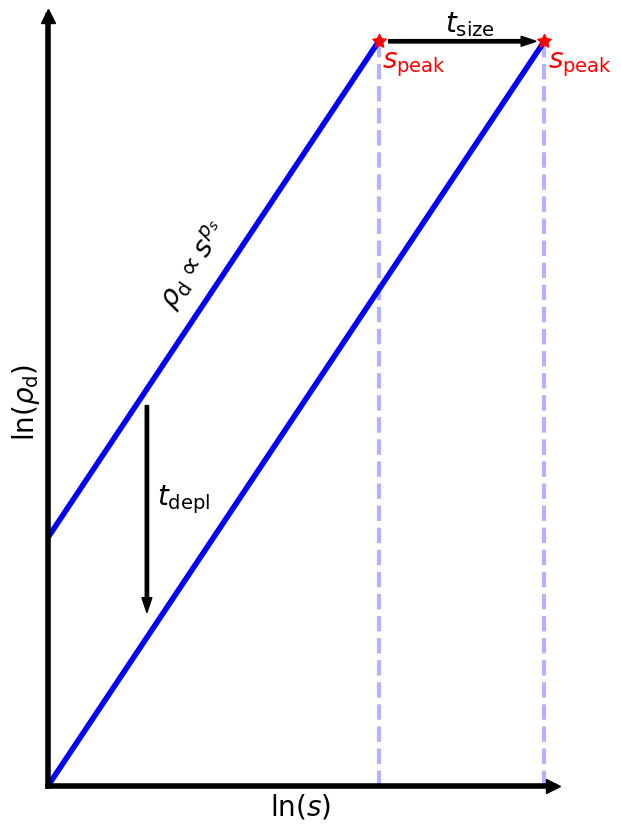}
    \caption{Schematic of the self-similar evolution of the grain mass density distribution as a function of size, showing the relation between the depletion of small grains (on a time scale $t_{\rm depl}$) due to growth to larger grains and the increase of grain size (on a timescale $t_{\rm size}$) also due to growth.} 
    \label{fig:tdepl_tsize}
\end{figure}

In the last subsection, we found that when the one-zone models run long enough, the effects of initial conditions wear off, with all solutions converging to $\rho_\mathrm{d, i} \propto s_i^{p_s}$, where $\rho_\mathrm{d, i}$ is the mass density per log size decade of the dust of size $s_i$ and $p_s\sim 1.9$ is the power-law index of the converged self-similar solution. This self-similar and attractive solution can also be described by a power law for the dust number density per (linear) size bin $n_i\propto s_i^{-q_s}$ and note that $p_s + q_s = 4$ by definition.\footnote{$\rho_\mathrm{d, i} = \frac{4}{3}\pi\tilde{\rho}_\mathrm{dm}s_i^3\Delta s_i n_i =  \frac{4}{3}\pi\tilde{\rho}_\mathrm{dm}s_i^3\tilde{A}s_i As_i^{-q} \propto s_i^{4-q_s} \propto s_i^{p_s}$, where $\Delta s_i = \tilde{A}s_i$, since our grain size bins are linear in log space. Thus $4 - q_s = p_s$.} For a semi-analytic treatment in this subsection, we will idealize the late-time dust mass density distribution shown in Fig.~\ref{fig:one_cell_q_and_chi} as a power-law with a sharp cut-off beyond the peak size $s_\mathrm{peak}$, as shown in Fig.~\ref{fig:tdepl_tsize}.  

To determine the ($e$-folding) time $t_{\rm size}$ for the peak grain size $s_\mathrm{peak}$ to increase by a factor of $e$ for the self-similar dust evolution sketched in Fig.~\ref{fig:tdepl_tsize}, we note that it is related to the ($e$-folding) time $t_{\rm depl}$ for the small grains at a fixed size to deplete by a factor of $e$ through
\begin{equation}
     t_{\rm size}\equiv \frac{dt}{d\ln s_{\rm peak}} = p_s\ t_{\rm depl}.
     \label{equ:tsize_tdepl}
\end{equation}
The depletion time $t_{\rm depl}$ for grains of a given size $s_k$ is given by 
\begin{equation}
     t_{\rm depl}\equiv \frac{dt} {d\ln \rho_{\rm d,k} }
     = \frac{dt}{d\ln n_k} =  - \eta\ f(s_k, s_s, s_l, q_s),
    \label{equ:dlns/dt}
\end{equation}
where equation~\ref{equ:N/dNdt} is used in the last equality, with the parameters $s_s$ and $s_l$ denoting, respectively, the lower and upper limit of the size integral in the expression for the function $f$ (see equation~\ref{equ:f(s_k; s_s, s_l, q)}). 

\begin{figure}
    \centering
    \includegraphics[width=\columnwidth]{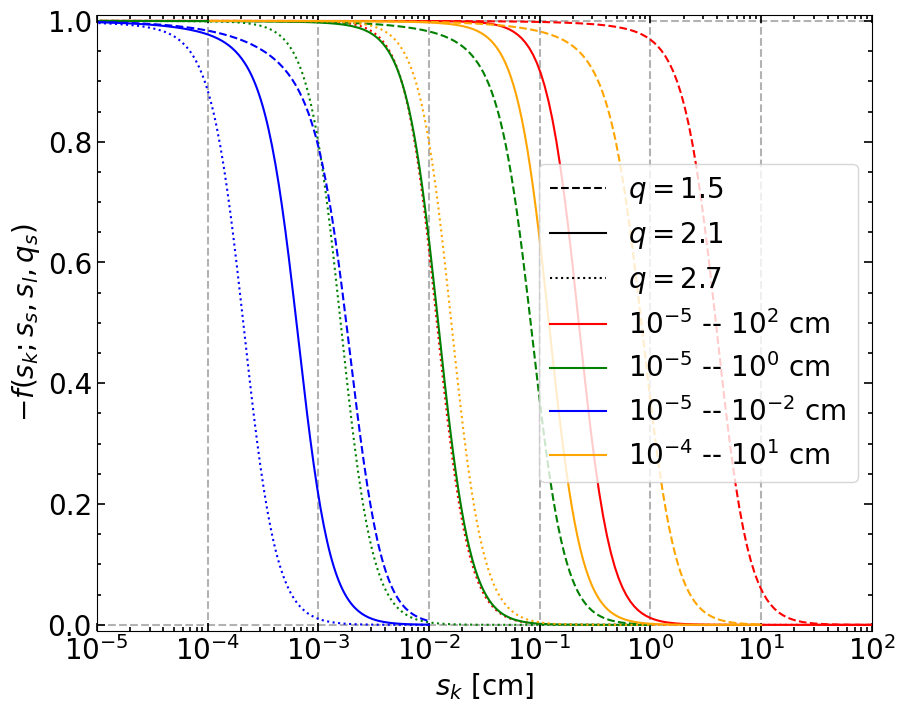}
    \caption{Values of $- f(s_k; s_s, s_l, q_s)$ (equation~\ref{equ:f(s_k; s_s, s_l, q)}). Each line is a different choice of parameters $s_s, s_l$ or $q_s$. Each linestyle represents a different $q_s$, with the dashed, solid and dotted lines have $q_s = 1.5, 2.1$ and $2.7$ respectively. Note that the solid lines with $q_s = 2.1$ fits the left half of self-similar and attractive solution in Fig.~\ref{fig:one_cell_q_and_chi} well. Each color denotes a different set of choice of $s_s$ and $s_l$, with values noted in the legend. The evalution of $f(s_k; s_s, s_l, q_s)$ is only done between $s_s$ and $s_l$. In all cases, if $s_k \ll s_l$, then $- f(s_k; s_s, s_l, q_s)\approx 1$ (the horizontal dashed gray line).}
    \label{fig:analytic_grain_growth_rate}
\end{figure}

We numerically integrate the function $-f(s_k; s_s, s_l, q_s)$ for four pairs of $s_s$ and $s_l$, and three values of $q_s$ (including the approximate value for the self-similar evolution, 2.1). The results are shown in Fig.~\ref{fig:analytic_grain_growth_rate}). For the purposes of comparison with the time scale for the peak grain size increase obtained numerically in the one-zone models (equation~\ref{equ:smax_fit_a}), the relevant values of the function $f$ are those for relatively small grains (much smaller than the maximum grain size, i.e., $s_k\ll s_l$) because the deviation of the size distribution near the maximum in the one-zone models is not captured by the strict power-law with a sharp cut-off adopted here. In the relevant limit $s_k\ll s_l$, the value of $-f(s_k; s_s, s_l, q_s)$ converges to $1$ independent of the choice of $s_s, s_l$ and $q_s$. Therefore, we have $t_{\rm depl}\approx \eta$ from equation~(\ref{equ:dlns/dt}) which, when plugged into  equation~(\ref{equ:tsize_tdepl}), yields 
\begin{equation}
     t_{\rm size} \approx p_s\ \eta = 1.9\ \eta = 2.5\ \frac{\rho_g\ v_\mathrm{th}}{\chi\ \rho_\mathrm{d, tot}\ \zeta},
     \label{equ:tsize_eta}
\end{equation}
which is remarkably close to the value of $1.84\ \eta$ numerically obtained from the one-zone models (see equation~\ref{equ:smax_fit_a}). We have used the definition of $\eta$ (equation~\ref{equ:smoluchowski_coag_only_define_eta}) in the last equality. 

 It is important to note that the time scale for the grain size increase $t_{\rm size}$ depends only on the local gas quantities through the quantity $\eta$. To give a feel for the physical value of $\eta$ (and thus $t_{\rm size}$), we scale the disk quantities to their values in the fiducial cell (cell \#1 in Fig.~\ref{fig:2D_eta}) 
\begin{equation}
    \begin{split}
        t_{\mathrm{size}} \approx 8680\ \mathrm{year}\  &\Big(\frac{0.01}{R_\mathrm{d-g}}\Big)
        \Big(\frac{v_\mathrm{th}}{10^5\ \mathrm{cm\ s^{-1}}}\Big) \\
        &
        \Big(\frac{2.5\cdot10^{-5}\ \mathrm{cm\ s^{-2}}}{\zeta}\Big)
        \Big(\frac{4}{\chi}\Big),
    \end{split}
    \label{equ:size-efolding-time}
\end{equation}
where $R_\mathrm{d-g} = \frac{\rho_\mathrm{d, tot}}{\rho_\mathrm{g}}$ is the local dust-to-gas ratio; $v_\mathrm{th}$ is the gas thermal speed, $\zeta$ is the dust acceleration from gas drag defined in equation~\ref{equ:delta_uij}. The time scale for the grain size to increase by one order of magnitude, $t_\mathrm{size-10}$, would be $\ln(10)$ times longer, so that 
\begin{equation}
    \begin{split}
        t_\mathrm{size-10} \approx 20000\ \mathrm{year}\ 
        &\Big(\frac{0.01}{R_\mathrm{d-g}}\Big)
        \Big(\frac{v_\mathrm{th}}{10^5\ \mathrm{cm\ s^{-1}}}\Big) \\
        &
        \Big(\frac{2.5\cdot10^{-5}\ \mathrm{cm\ s^{-2}}}{\zeta}\Big)
        \Big(\frac{4}{\chi}\Big).
    \end{split}
    \label{equ:1-order_time}
\end{equation} 
This estimate holds as long as the grain-grain collision speed is given by $\Delta u_{ij} = \chi \zeta|t_{s_i} - t_{s_j}|$. The typical values of $t_\mathrm{size}$ are estimated in the discussion section (\S~\ref{sec:discussion}) below. 

\subsection{Physical Interpretation of \texorpdfstring{$\eta$}{eta}}
\label{sec:eta}

As discussed above, the quantity $\eta$ defined in equation~\ref{equ:smoluchowski_coag_only_define_eta} is a crucial parameter for grain growth: it is about half of the time-scale $t_\mathrm{size}$ for the maximum grain size to increase by a factor of $e$ (see equation~\ref{equ:tsize_eta}). The physical meaning of this parameter can be obtained from the following consideration. 

Since the grain mass distribution typically peaks around some size $s_{\rm peak}$ (see Figs.~\ref{fig:one_cell_q_and_chi} and \ref{fig:tdepl_tsize}), we can obtain a rough estimate of the number density $n_{\rm peak}$ of the grains with the peak size by assuming all of the dust mass is contained in such grains:
\begin{equation}
    n_{\rm peak} \approx \frac{3\rho_\mathrm{d,tot}}{4\pi s_{\rm peak}^3
    \tilde{\rho}_\mathrm{dm} },
    \label{equ:n_peak}
\end{equation}
where $\rho_\mathrm{d,tot}$ is the total grain mass density and $\tilde{\rho}_\mathrm{dm}$ is the dust material density. This, combined with the cross-section $\pi s_{\rm peak}^2$, yields the mean-free-path for a collision with grains of size $s_{\rm peak}$:
\begin{equation}
    l_{\rm peak}=\frac{1}{\pi s_{\rm peak}^2 n_{\rm peak}}\approx \frac{4\tilde{\rho}_\mathrm{dm} s_{\rm peak}}{3\rho_\mathrm{d,tot}}.
    \label{equ:mfp}
\end{equation}
Since the grains of the peak size drift relative to gas at a terminal speed of 
\begin{equation}
    u_{\rm d,peak}= \chi \zeta \frac{\tilde{\rho}_\mathrm{dm}\ s_{\rm peak}}{\rho_\mathrm{g}\ u_\mathrm{th}},
    \label{equ:peak_drift}
\end{equation}
we have from the definition of $\eta$ (equation~\ref{equ:smoluchowski_coag_only_define_eta}) that
\begin{equation}
    \eta \approx \frac{l_{\rm peak}}{u_{\rm d,peak}}.
    \label{equ:eta_meaning}
\end{equation}
In other words, the parameter $\eta$ is basically the time it takes a grain near the peak size (where most of the dust mass resides) to collide with another grain of comparable size, which sets the timescale for grain growth. It is basically the same as the characteristic time of  grain-grain collision estimated by \citet[][see their equation (9)]{Nakagawa1981}.

\section{Discussion}
\label{sec:discussion}

\subsection{Validity of the terminal velocity approximation}
\label{sec:terminal_vel_and_limitation}

\begin{figure*}
    \centering
    \includegraphics[width=\textwidth]{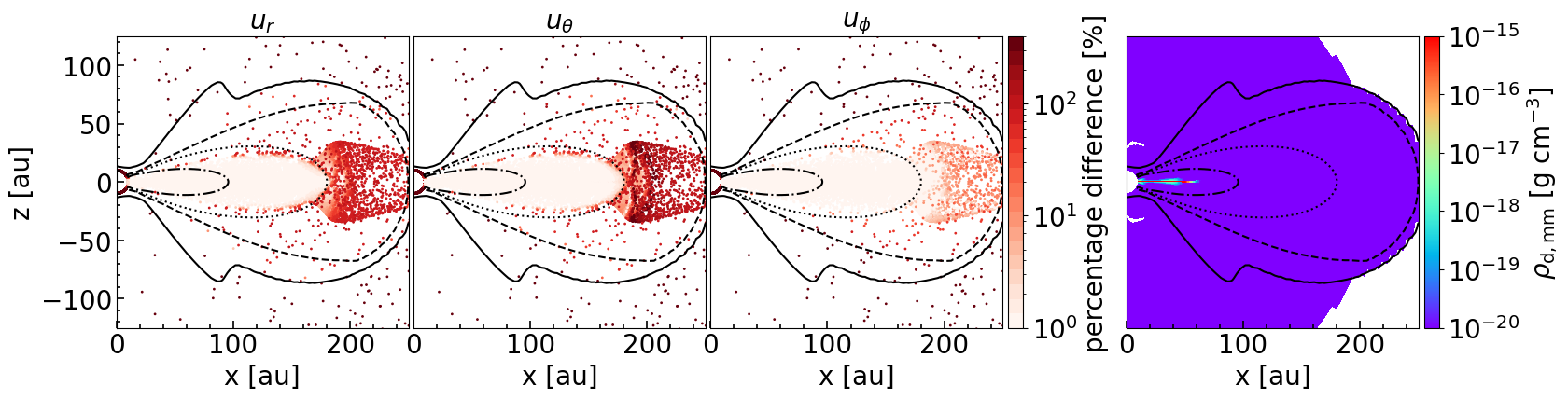}
    \caption{{\bf Left three panels}: Percentage difference between the terminal velocities used in the Eulerian simulation and those calculated in the Lagrangian simulation for mm-sized grains. The black lines are iso-density contours that are the same in each plot. {\bf Right}: the spatial distribution of the mm-sized grains in the Eulerian simulation with $\chi=4$, showing that such large grains are concentrated in the high density regions where the terminal velocity approximation is good to the percent-level.}
    \label{fig:terminal_vel_1e-1} 
\end{figure*}

The grain velocities used in the Eulerian models are the ``terminal velocities'' (equation~\ref{equ:vel_terminal}) of the grains at their respective locations. To test their validity, we compare them to those obtained from the Lagrangian simulation, where the particle velocities are calculated self-consistently, as described in \S~\ref{sec:dust_Lagrangian_method}. Because the Lagrangian particles are usually not located at cell centers where the gas quantities in the (Eulerian) hydro simulation are defined, we use the TSC interpolation algorithm (see section \S~\ref{sec:dust_Lagrangian_method}) to obtain the gas quantities at the grain locations.  Fig.~\ref{fig:terminal_vel_1e-1} shows the percentage difference between the velocities in the Eulerian and Lagrangian models for mm-sized grains (the left three panels), as well as the spatial distribution of such grains in the $\chi = 4$ Eulerian model (the rightmost panel). We see that most mm-sized grains in the Eulerian model are concentrated within the dashed-dotted line (where the gas density is greater than $4\cdot10^{-14}$g~cm$^{-2}$), where the difference is $<1\%$. A larger difference exists in the lower density regions (particularly beyond the dotted line marking $4\cdot10^{-15}$g~cm$^{-2}$) where mm-sized grains coming from the envelope in one hemisphere can pass through the disk midplane to the other hemisphere before settling back to the midplane \citep[see, e.g.][]{Bate2016}. Such a behavior cannot be captured by the Eulerian model but there are very few mm-sized grains in the envelope and the low-density part of the disk, so the difference is inconsequential as far as grain growth is concerned. The situation is similar for the cm-sized grains, which are concentrated in even higher density regions. The terminal velocity approximation is even better for the smaller grains, which are better coupled to the gas. 

 \subsection{Slow Grain Growth in a Laminar Protostellar Disk}
\label{sub:slow_growth}
%
%
%
%

In a laminar disk without any turbulence, the main driver for grain growth is the force that causes grains of different sizes to move at different (terminal) speeds. In the vertical direction, this force is the vertical component of the gravity acting on the grain that is balanced by the aerodynamic drag. At a given cylindrical radius $R$, this force points towards the disk midplane, and increases roughly linearly with the distance from the midplane. Its characteristic value can be estimated at one disk scale height $h=R\ {c_s}/{V_K}$: 
\begin{equation}
    g_{\rm z,h}\approx \frac{GM_*}{R^2}\frac{h}{R}\approx \Omega_K c_s
    \label{equ:g_zh}
\end{equation}
where $c_s = \sqrt{\frac{k_b T}{\mu m_H}}$ is the isothermal sound speed, $V_K$ the local Keplerian speed, and $\Omega_K=V_K/R$ is the orbital frequency. 

Using $g_{\rm z,h}$ to approximate the dust acceleration to be balanced by the aerodynamic drag, $\zeta$, we obtain an estimate for the grain size $e$-folding time at one scale height:
\begin{equation}
\begin{split}
    & t_{\rm size,h}\approx 1.9\ \eta \
    = 2.53 \frac{1}{\chi R_\mathrm{d-g}} \frac{v_\mathrm{th}}{g_{\rm z,h}}\approx \frac{64}{\chi}\left(\frac{0.01}{R_\mathrm{d-g}}\right)\ P_K \\
    & = 90,000\ ({\rm years})\  \left(\frac{1}{\chi}\right)\left(\frac{0.01}{R_\mathrm{d-g}}\right)\ \left(\frac{0.5 M_\odot}{M_*}\right)^{1/2} \left(\frac{R}{100\ {\rm au}}\right)^{3/2}, 
    \end{split}
   \label{equ:tsize_h} 
\end{equation}
where $P_K=2\pi/\Omega_K$ is the local orbital period. Note that this $e$-folding time is similar to (but $\sim 2.5$ times longer than) the estimate of grain growth time given in \citet[][their equation (23)]{Tanaka2005}, which was taken from \citet[][]{Nakagawa1981} but with a factor of $4/3$ removed (i.e., the value adopted in \citet[][]{Tanaka2005} is only $75\%$ of that of \citet[][]{Nakagawa1981}).

We have checked the above estimate against the baseline model without any enhancement (i.e., $\chi=1$) and found that the above estimate is about $50\%$ longer than that obtained numerically in the model. The reason for the discrepancy is that the protostellar disk formed in our simulation is relatively massive, with a significant self-gravity in the vertical direction that is about half of that from the central star at one scale height; the disk self-gravity increases the dust settling (and thus grain-grain collision) speed somewhat, with a corresponding reduction in the grain growth time. Nevertheless, the above equation provides a reasonable estimate for the grain size $e$-folding timescale. 

Equation~(\ref{equ:tsize_h}) highlights the fundamental difficulty with grain growth from micron-sizes to mm-sizes in large (100~au-sized), laminar protostellar disks revealed by our baseline model (see Fig.~\ref{fig:74_34-74_32_summary_plot}). Increasing the grain size by a factor of $10^3$ (from 1~$\mu$m to 1~mm) requires 6.9 $e$-folding times, which corresponds to $\sim 6.2\times 10^5$~years for the fiducial stellar mass and radius adopted in equation~(\ref{equ:tsize_h}). It is significantly longer than the time scale for the Class 0 stage of star formation of $\sim 1.6\times 10^5$~years estimated by \citep{Evans2009}. If we choose a smaller characteristic height than the pressure scale height $h$ to evaluate the net dust acceleration to be balanced by the drag acceleration ($\zeta$), as may be appropriate for large grains, the grain size $e-$folding time would be even longer because of a smaller local vertical gravity that would drive a slower grain-grain collision. The longer size $e-$folding time would strengthen our conclusion, namely, in order to produce a substantial amount of mm-sized grains in the relatively short deeply embedded Class 0 phase of star formation, the grain-grain collision speeds must be enhanced over those of the baseline laminar case. 





\subsection{Enhanced Growth Through Higher Grain-Grain Collision Speeds}
\label{sub:enhanced}

To illustrate how a higher grain-grain collision speed enhances the grain growth, we carried out a simulation with a factor of $\chi=4$ higher speed in \S~\ref{sec:dust_the_chi=4_model}. We found that, by the time $t=88,750$ years (or 48,750 years after the formation of the disk and the grain growth in it, when the disk radius reaches $250$~au), about 140~$M_\oplus$ of grains have grown beyond $0.1$~mm  and about 15~$M_\oplus$ beyond 1~mm (see Fig.~\ref{fig:74_34-74_32_summary_plot}). 

Part of the reason for the substantial grain growth in this case can be understood from the size $e$-folding time $t_{\rm size,h}$ at one scale height (equation~\ref{equ:tsize_h}). For an enhancement factor of $\chi=4$, $t_{\rm size,h}$ is reduced to about 22,500~years at 100~au, which is well within the timescale for the disk to form and grow. This is reflected, for example, in Fig.~\ref{fig:vertical_grain_distribution}, which shows the progression of the grain size increase from the disk upper surface towards the midplane at a representative cylindrical radius $50$~au, where $t_{\rm size,h}$ is shorter still (about 8,000~years). The grain size peaks around $40~\mu$m at a height of $\sim 10$~au. Most of the grains grown at relatively large vertical heights are quickly advected towards the midplane (see  Fig.~\ref{fig:74_10_dust_disk_rho} for an example of advection timescale), where they grow further. Further growth near the midplane may appear surprising at the first sight, because the vertical component of the gravity $g_z$, which drives the grain growth through dust-gas drift at relatively large vertical heights, disappears on the midplane. 
%
%
%
%
%

However, near the midplane, the grains can drift radially inward relative to the gas because of the gas pressure gradient. We can estimate the acceleration due to this gradient as
\begin{equation}
    a_p=\left| \frac{1}{\rho} \frac{d\ P}{d\ R}\right| \approx \alpha_p \frac{c_s^2}{R} \approx \alpha_p \frac{c_s}{V_K} g_{\rm z,h}\approx \alpha_p \frac{h}{R} g_{\rm z,h},
    \label{equ:a_p}
\end{equation}
where $\alpha_p$ is the exponent in the pressure distribution $P\propto R^{-\alpha_p}$, $R$ the cylindrical radius, $c_s=\sqrt{P/\rho}$ the isothermal sound, $V_K$ the Keplerian speed, and $g_{\rm z,h}\approx \Omega_K\ c_s$ the vertical component of the stellar gravity at one scale height (see equation~\ref{equ:g_zh}). 
Using $a_p$ to approximate the dust acceleration from aerodynamic drag, $\zeta$, we obtain an estimate for the grain size $e$-folding time on the disk midplane: 
\begin{equation}
\begin{split}
    & t_{\rm size,mid}\approx 1.9\ \eta \
    = 2.53 \frac{1}{\chi R_\mathrm{d-g}} \frac{v_\mathrm{th}}{a_p}\approx  \frac{64}{\chi}\left(\frac{0.01}{R_\mathrm{d-g}}\right)\ \frac{1}{\alpha_p}\ \frac{1}{h/R}\ P_K \\
    & = 3\times 10^5 ({\rm yrs}) \left(\frac{1}{\chi}\right)\left(\frac{0.01}{R_\mathrm{d-g}}\right)\left(\frac{3}{\alpha_p}\right)\left(\frac{0.1}{h/R}\right)\left(\frac{0.5 M_\odot}{M_*}\right)^{1/2} \left(\frac{R}{10^2{\rm au}}\right)^{3/2}. 
    \end{split}
   \label{equ:tsize_mid} 
\end{equation}

Note that the force due to the radial pressure gradient on the midplane is smaller than the vertical gravity at one scale height by a factor of $\sim \alpha_p\ h/R$ (or $\sim 0.3$ for the adopted fiducial values of $\alpha_p=3$ and $h/R=0.1$), which translates to a longer size $e$-folding time on the midplane (by a factor of $\sim 10/3$) compared to that at one scale height, with everything else being equal. However, the advection of dust grains towards the midplane makes the midplane dust-to-gas ratio $R_{\rm d-g}$ substantially higher than the initial value of 0.01 in the envelope, reaching $R_{\rm d-g}\approx 0.03$ near the inner edge of the disk and at a radius $R\approx 50$~au (see Fig.~\ref{fig:dust_to_gas_ratio}). The higher dust-to-gas ratio leads to a shorter grain growth time scale. 

For example, with $R_{\rm d-g}\approx 0.03$ and $\chi=4$, the size $e$-folding time at a radius around $50$~au on the midplane becomes $t_{\rm size,mid}\approx 8,840$~years if other parameters are held at their fiducial values. This is significantly shorter than the disk formation and evolution time during the Class 0 phase. The relatively short size $e$-folding timescale is consistent with the fact that, at this midplane location, the majority of the grains have already grown to a size of order $300~\mu$m by the time shown in Fig.~\ref{fig:vertical_grain_distribution} (46,250 years after the disk formation and the start of grain growth; see the right-most curve). 

A salient feature of the $\chi=4$ model with an enhanced grain-grain collision speed is that grain growth in the atmosphere of the disk at relatively large heights can lead to a faster settling of the grown dust to the midplane, which increases the local dust-to-gas ratio, which, in turn, speeds up further growth there. Whether such an enhancement of collision speed can be achieved in protostellar disks is unclear. We believe that it is plausible during the Class 0 stage of star formation, when the protostar is expected to accrete most of its final mass. \citet{Yen2017} estimated a mass accretion rate of order $10^{-6}$ -- $10^{-5}$ M$_\odot$~yr$^{-1}$, which is 2 to 3 orders of magnitude higher than the typical value for classical T Tauri stars. What drives the fast accretion in Class 0 disks is still unclear. If it is due to turbulent motions \citep[from, e.g., magneto-rotational instability;][]{Balbus1991}, the required turbulent speed would be comparable to the sound speed (i.e., with an effective $\alpha$-parameter of order unity as opposed to the typical value of $10^{-2}$ for classical T Tauri disks). It is well known that turbulence can in principle promote grain growth by increasing the grain-grain collision speed \citep{Voelk1980, MMV1991, OrmelCuzzi2007}. However, turbulence can also puff up the vertical dust distribution, which can make the grain growth more difficult by reducing the dust number density, particularly for relatively large grains \citep{Ohashi2021}. The degree of enhancement depends on the details of the turbulence in the Class 0 disks, which are currently unknown. Our results highlight the pressing need to tackle the problem of enhanced grain growth through disk turbulence or other means during the earliest deeply embedded phase of star formation. 

\section{Conclusion}
\label{sec:conclusion}

We have studied the growth of dust grains during the formation of protostellar disks. We follow the disk formation out of the gravitational collapse of a rotating molecular cloud core using 2D (axisymmetric) radiation hydrodynamic simulations, which form the basis for the grain growth calculations that are performed as a post-processing step. Our main conclusions are as follows:

1. We find that the so-called ``terminal velocity approximation" (equation~\ref{equ:vel_terminal}) holds in most of the laminar protostellar disks, particularly the relatively dense regions that are of the most interest to grain growth (see Fig.~\ref{fig:terminal_vel_1e-1}). The approximation greatly simplifies the calculations of the dust drift speed relative to the gas and thus the collision speeds between grains of different sizes, which control the rate of grain growth in our simulations.

2. We find numerically that the grain-grain collision from differential (size-dependent) terminal velocities alone is too slow to convert a significant fraction of the grains into mm/cm sizes during the deeply embedded Class 0 phase (\S~\ref{sec:dust_the_fiducial_model}). This difficulty is caused by a rather long grain size $e$-folding time, which is estimated to be nearly two orders of magnitude longer than the local orbital period (or $\sim 10^5$~years on the 100~au scale; see equation~(\ref{equ:tsize_h}) and the need for several $e$-folding times to grow from $\mu$m to mm sizes). Enhanced grain-grain collision speeds are required to produce enough mm/cm sized grains for, e.g., giant planet formation through core accretion in the Class 0 phase. 

3. We find substantial grain growth when the grain-grain collision speed is enhanced by a factor of 4 (\S~\ref{sec:dust_the_chi=4_model}), with $\sim 15\ M_\oplus$ of grains larger than 1 mm and $\sim 140\ M_\oplus$ larger than 0.1~mm when the disk size reaches 250~au (see Fig.~\ref{fig:74_34-74_32_summary_plot}). As the grains grow bigger in the atmosphere of the disk, they settle faster towards the midplane, which increases the local dust-to-gas ratio, which, in turn, speeds up further growth there. Most of the largest, mm/cm-sized grains are formed near the midplane in the inner part of the disk (see Fig.~\ref{fig:74_10_dust_disk_rho}).  

4. If a substantial amount of mm/cm sized grains exist in Class 0 disks, as indicated by thermal dust emission at cm wavelengths and required if planets are to form in this earliest phase of star formation, the grain-grain collision speeds must be increased well beyond that given by the differential (size-dependent) terminal velocities. How this enhancement is achieved is unclear, although turbulence is a strong possibility. Our study highlights the urgent need for quantifying the level of turbulence in Class 0 disks and its effect on grain growth. 
\section*{Acknowledgements}

We thank Hiroyuki Hirashita, Zhaohuan Zhu and Jon Ramsey for useful discussions and the referee, H. Kobayashi, for constructive comments that improved the presentation of the paper. YT acknowledges support from an interdisciplinary fellowship and a computing allocation from the University of Virginia, NSF AST-1815784, and a NASA High-End Computing allocation. ZYL is supported in part by NASA 80NSSC20K0533 and NSF AST-1716259. KHL acknowledges support from SOFIA grant 07-0235. 

\section*{Data Availability}
The data underlying this article will be shared on reasonable request to the corresponding author.




\bibliographystyle{mnras}
\bibliography{biblio} 

\begin{thebibliography}{}
\makeatletter
\relax
\def\mn@urlcharsother{\let\do\@makeother \do\$\do\&\do\#\do\^\do\_\do\%\do\~}
\def\mn@doi{\begingroup\mn@urlcharsother \@ifnextchar [ {\mn@doi@}
  {\mn@doi@[]}}
\def\mn@doi@[#1]#2{\def\@tempa{#1}\ifx\@tempa\@empty \href
  {http://dx.doi.org/#2} {doi:#2}\else \href {http://dx.doi.org/#2} {#1}\fi
  \endgroup}
\def\mn@eprint#1#2{\mn@eprint@#1:#2::\@nil}
\def\mn@eprint@arXiv#1{\href {http://arxiv.org/abs/#1} {{\tt arXiv:#1}}}
\def\mn@eprint@dblp#1{\href {http://dblp.uni-trier.de/rec/bibtex/#1.xml}
  {dblp:#1}}
\def\mn@eprint@#1:#2:#3:#4\@nil{\def\@tempa {#1}\def\@tempb {#2}\def\@tempc
  {#3}\ifx \@tempc \@empty \let \@tempc \@tempb \let \@tempb \@tempa \fi \ifx
  \@tempb \@empty \def\@tempb {arXiv}\fi \@ifundefined
  {mn@eprint@\@tempb}{\@tempb:\@tempc}{\expandafter \expandafter \csname
  mn@eprint@\@tempb\endcsname \expandafter{\@tempc}}}

\bibitem[\protect\citeauthoryear{Acheson \& Acheson}{Acheson \&
  Acheson}{1990}]{acheson1990elementary}
Acheson D.,  Acheson F.,  1990, Elementary Fluid Dynamics.
Comparative Pathobiology - Studies in the Postmodern Theory of Education,
  Clarendon Press, \url {https://books.google.com/books?id=GgC69-WUTs0C}

\bibitem[\protect\citeauthoryear{{Andre}, {Ward-Thompson}  \&
  {Barsony}}{{Andre} et~al.}{1993}]{Andre1993}
{Andre} P.,  {Ward-Thompson} D.,   {Barsony} M.,  1993, \mn@doi [\apj]
  {10.1086/172425}, \href
  {https://ui.adsabs.harvard.edu/abs/1993ApJ...406..122A} {406, 122}

\bibitem[\protect\citeauthoryear{Andrews}{Andrews}{2020}]{Andrews2020}
Andrews S.~M.,  2020, \mn@doi [Annual Review of Astronomy and Astrophysics]
  {10.1146/annurev-astro-031220-010302}, 58, 483

\bibitem[\protect\citeauthoryear{Andrews et~al.,}{Andrews
  et~al.}{2018}]{Andrews2018}
Andrews S.~M.,  et~al., 2018, \mn@doi [The Astrophysical Journal]
  {10.3847/2041-8213/aaf741}, 869, L41

\bibitem[\protect\citeauthoryear{{Armitage}}{{Armitage}}{2015}]{Armitage2015}
{Armitage} P.~J.,  2015, arXiv e-prints, \href
  {https://ui.adsabs.harvard.edu/abs/2015arXiv150906382A} {p. arXiv:1509.06382}

\bibitem[\protect\citeauthoryear{{Balbus} \& {Hawley}}{{Balbus} \&
  {Hawley}}{1991}]{Balbus1991}
{Balbus} S.~A.,  {Hawley} J.~F.,  1991, \mn@doi [\apj] {10.1086/170270}, \href
  {https://ui.adsabs.harvard.edu/abs/1991ApJ...376..214B} {376, 214}

\bibitem[\protect\citeauthoryear{Bate \& Lorén-Aguilar}{Bate \&
  Lorén-Aguilar}{2016}]{Bate2016}
Bate M.~R.,  Lorén-Aguilar P.,  2016, \mn@doi [Monthly Notices of the Royal
  Astronomical Society] {10.1093/mnras/stw2853}, 465, 1089

\bibitem[\protect\citeauthoryear{{Beckwith}}{{Beckwith}}{1999}]{Beckwith1999}
{Beckwith} S. V.~W.,  1999, in {Lada} C.~J.,  {Kylafis} N.~D.,  eds,  NATO
  Advanced Study Institute (ASI) Series C Vol. 540, The Origin of Stars and
  Planetary Systems. p.~579 (\mn@eprint {arXiv} {astro-ph/9905003})

\bibitem[\protect\citeauthoryear{{Birnstiel}, {Dullemond}  \&
  {Brauer}}{{Birnstiel} et~al.}{2010}]{Birnstiel2010}
{Birnstiel} T.,  {Dullemond} C.~P.,   {Brauer} F.,  2010, \mn@doi [\aap]
  {10.1051/0004-6361/200913731}, \href
  {https://ui.adsabs.harvard.edu/abs/2010A&A...513A..79B} {513, A79}

\bibitem[\protect\citeauthoryear{{Chandler}, {Barsony}  \& {Moore}}{{Chandler}
  et~al.}{1998}]{Chandler1998}
{Chandler} C.~J.,  {Barsony} M.,   {Moore} T. J.~T.,  1998, \mn@doi [\mnras]
  {10.1046/j.1365-8711.1998.01818.x}, \href
  {https://ui.adsabs.harvard.edu/abs/1998MNRAS.299..789C} {299, 789}

\bibitem[\protect\citeauthoryear{{Chang}, {Davis}  \& {Jiang}}{{Chang}
  et~al.}{2020}]{Chang_et_al.2020}
{Chang} P.,  {Davis} S.~W.,   {Jiang} Y.-F.,  2020, \mn@doi [\mnras]
  {10.1093/mnras/staa573}, \href
  {https://ui.adsabs.harvard.edu/abs/2020MNRAS.493.5397C} {493, 5397}

\bibitem[\protect\citeauthoryear{{Cridland}, {Rosotti}, {Tabone}, {Tychoniec},
  {McClure}  \& {van Dishoeck}}{{Cridland} et~al.}{2021}]{CrIdland2021}
{Cridland} A.~J.,  {Rosotti} G.~P.,  {Tabone} B.,  {Tychoniec} L.,  {McClure}
  M.,   {van Dishoeck} E.~F.,  2021, arXiv e-prints, \href
  {https://ui.adsabs.harvard.edu/abs/2021arXiv211206734C} {p. arXiv:2112.06734}

\bibitem[\protect\citeauthoryear{{Draine}}{{Draine}}{2006}]{Draine2006}
{Draine} B.~T.,  2006, \mn@doi [\apj] {10.1086/498130}, \href
  {https://ui.adsabs.harvard.edu/abs/2006ApJ...636.1114D} {636, 1114}

\bibitem[\protect\citeauthoryear{Epstein}{Epstein}{1924}]{Epstein1924}
Epstein P.~S.,  1924, \mn@doi [Phys. Rev.] {10.1103/PhysRev.23.710}, 23, 710

\bibitem[\protect\citeauthoryear{{Evans} Neal~J. et~al.,}{{Evans}
  et~al.}{2009}]{Evans2009}
{Evans} Neal~J. I.,  et~al., 2009, \mn@doi [\apjs]
  {10.1088/0067-0049/181/2/321}, \href
  {https://ui.adsabs.harvard.edu/abs/2009ApJS..181..321E} {181, 321}

\bibitem[\protect\citeauthoryear{{Fedele, D.} et~al.,}{{Fedele, D.}
  et~al.}{2018}]{Fedele2018}
{Fedele, D.} et~al., 2018, \mn@doi [A\&A] {10.1051/0004-6361/201731978}, 610,
  A24

\bibitem[\protect\citeauthoryear{{Galametz}, {Maury}, {Valdivia}, {Testi},
  {Belloche}  \& {Andr{\'e}}}{{Galametz} et~al.}{2019}]{Galametz2019}
{Galametz} M.,  {Maury} A.~J.,  {Valdivia} V.,  {Testi} L.,  {Belloche} A.,
  {Andr{\'e}} P.,  2019, \mn@doi [\aap] {10.1051/0004-6361/201936342}, \href
  {https://ui.adsabs.harvard.edu/abs/2019A&A...632A...5G} {632, A5}

\bibitem[\protect\citeauthoryear{{Haffert}, {Bohn}, {de Boer}, {Snellen},
  {Brinchmann}, {Girard}, {Keller}  \& {Bacon}}{{Haffert}
  et~al.}{2019}]{Haffert2019}
{Haffert} S.~Y.,  {Bohn} A.~J.,  {de Boer} J.,  {Snellen} I.~A.~G.,
  {Brinchmann} J.,  {Girard} J.~H.,  {Keller} C.~U.,   {Bacon} R.,  2019,
  \mn@doi [Nature Astronomy] {10.1038/s41550-019-0780-5}, \href
  {https://ui.adsabs.harvard.edu/abs/2019NatAs...3..749H} {3, 749}

\bibitem[\protect\citeauthoryear{{Hasegawa}, {Suzuki}, {Tanaka}, {Kobayashi}
  \& {Wada}}{{Hasegawa} et~al.}{2021}]{Hasegawa2021}
{Hasegawa} Y.,  {Suzuki} T.~K.,  {Tanaka} H.,  {Kobayashi} H.,   {Wada} K.,
  2021, \mn@doi [\apj] {10.3847/1538-4357/abf6cf}, \href
  {https://ui.adsabs.harvard.edu/abs/2021ApJ...915...22H} {915, 22}

\bibitem[\protect\citeauthoryear{{Hirashita} \& {Li}}{{Hirashita} \&
  {Li}}{2013}]{Hirashita&Li2013}
{Hirashita} H.,  {Li} Z.~Y.,  2013, \mn@doi [\mnras] {10.1093/mnrasl/slt081},
  \href {https://ui.adsabs.harvard.edu/abs/2013MNRAS.434L..70H} {434, L70}

\bibitem[\protect\citeauthoryear{{Hockney} \& {Eastwood}}{{Hockney} \&
  {Eastwood}}{1981}]{TSC}
{Hockney} R.~W.,  {Eastwood} J.~W.,  1981, {Computer Simulation Using
  Particles}.
McGraw-Hill International Book Company

\bibitem[\protect\citeauthoryear{{Hosokawa} \& {Omukai}}{{Hosokawa} \&
  {Omukai}}{2009}]{Hosokawa&Omukai2009}
{Hosokawa} T.,  {Omukai} K.,  2009, \mn@doi [\apj]
  {10.1088/0004-637X/691/1/823}, \href
  {https://ui.adsabs.harvard.edu/abs/2009ApJ...691..823H} {691, 823}

\bibitem[\protect\citeauthoryear{Isella et~al.,}{Isella
  et~al.}{2016}]{Isella2016}
Isella A.,  et~al., 2016, \mn@doi [Phys. Rev. Lett.]
  {10.1103/PhysRevLett.117.251101}, 117, 251101

\bibitem[\protect\citeauthoryear{{J{\o}rgensen} et~al.,}{{J{\o}rgensen}
  et~al.}{2007}]{Jorgensen2007}
{J{\o}rgensen} J.~K.,  et~al., 2007, \mn@doi [\apj] {10.1086/512230}, \href
  {https://ui.adsabs.harvard.edu/abs/2007ApJ...659..479J} {659, 479}

\bibitem[\protect\citeauthoryear{{Kobayashi} \& {Tanaka}}{{Kobayashi} \&
  {Tanaka}}{2010}]{Kobayashi2010}
{Kobayashi} H.,  {Tanaka} H.,  2010, \mn@doi [\icarus]
  {10.1016/j.icarus.2009.10.004}, \href
  {https://ui.adsabs.harvard.edu/abs/2010Icar..206..735K} {206, 735}

\bibitem[\protect\citeauthoryear{{Kobayashi} \& {Tanaka}}{{Kobayashi} \&
  {Tanaka}}{2021}]{Kobayashi2021}
{Kobayashi} H.,  {Tanaka} H.,  2021, \mn@doi [\apj] {10.3847/1538-4357/ac289c},
  \href {https://ui.adsabs.harvard.edu/abs/2021ApJ...922...16K} {922, 16}

\bibitem[\protect\citeauthoryear{{Kobayashi}, {Tanaka}  \&
  {Okuzumi}}{{Kobayashi} et~al.}{2016}]{Kobayashi2016}
{Kobayashi} H.,  {Tanaka} H.,   {Okuzumi} S.,  2016, \mn@doi [\apj]
  {10.3847/0004-637X/817/2/105}, \href
  {https://ui.adsabs.harvard.edu/abs/2016ApJ...817..105K} {817, 105}

\bibitem[\protect\citeauthoryear{{Kuiper}, {Klahr}, {Beuther}  \&
  {Henning}}{{Kuiper} et~al.}{2010}]{Kuiper_et_al._2010}
{Kuiper} R.,  {Klahr} H.,  {Beuther} H.,   {Henning} T.,  2010, \mn@doi [\apj]
  {10.1088/0004-637X/722/2/1556}, \href
  {https://ui.adsabs.harvard.edu/abs/2010ApJ...722.1556K} {722, 1556}

\bibitem[\protect\citeauthoryear{Kwon, Looney, Mundy  \& Welch}{Kwon
  et~al.}{2015}]{Kwon2015}
Kwon W.,  Looney L.~W.,  Mundy L.~G.,   Welch W.~J.,  2015, \mn@doi [The
  Astrophysical Journal] {10.1088/0004-637x/808/1/102}, 808, 102

\bibitem[\protect\citeauthoryear{{Lada}}{{Lada}}{1987}]{Lada1987}
{Lada} C.~J.,  1987, in {Peimbert} M.,  {Jugaku} J.,  eds,  IAU Symposium Vol.
  115, Star Forming Regions. p.~1

\bibitem[\protect\citeauthoryear{{Laor} \& {Draine}}{{Laor} \&
  {Draine}}{1993}]{Laor&Draine1993}
{Laor} A.,  {Draine} B.~T.,  1993, \mn@doi [\apj] {10.1086/172149}, \href
  {https://ui.adsabs.harvard.edu/abs/1993ApJ...402..441L} {402, 441}

\bibitem[\protect\citeauthoryear{{Li}, {Banerjee}, {Pudritz}, {J{\o}rgensen},
  {Shang}, {Krasnopolsky}  \& {Maury}}{{Li} et~al.}{2014}]{Li2014}
{Li} Z.~Y.,  {Banerjee} R.,  {Pudritz} R.~E.,  {J{\o}rgensen} J.~K.,  {Shang}
  H.,  {Krasnopolsky} R.,   {Maury} A.,  2014, in {Beuther} H.,  {Klessen}
  R.~S.,  {Dullemond} C.~P.,   {Henning} T.,  eds, Protostars and Planets VI.
  p.~173 (\mn@eprint {arXiv} {1401.2219}),
  \mn@doi{10.2458/azu\_uapress\_9780816531240-ch008}

\bibitem[\protect\citeauthoryear{{Lombart} \& {Laibe}}{{Lombart} \&
  {Laibe}}{2021}]{Lombart2021}
{Lombart} M.,  {Laibe} G.,  2021, \mn@doi [\mnras] {10.1093/mnras/staa3682},
  \href {https://ui.adsabs.harvard.edu/abs/2021MNRAS.501.4298L} {501, 4298}

\bibitem[\protect\citeauthoryear{{Markiewicz}, {Mizuno}  \&
  {Voelk}}{{Markiewicz} et~al.}{1991}]{MMV1991}
{Markiewicz} W.~J.,  {Mizuno} H.,   {Voelk} H.~J.,  1991, \aap, \href
  {https://ui.adsabs.harvard.edu/abs/1991A&A...242..286M} {242, 286}

\bibitem[\protect\citeauthoryear{{Mathis}, {Rumpl}  \& {Nordsieck}}{{Mathis}
  et~al.}{1977}]{Mathis_et_al.1977}
{Mathis} J.~S.,  {Rumpl} W.,   {Nordsieck} K.~H.,  1977, \mn@doi [\apj]
  {10.1086/155591}, \href
  {https://ui.adsabs.harvard.edu/abs/1977ApJ...217..425M} {217, 425}

\bibitem[\protect\citeauthoryear{{Morfill}, {Roeser}, {Voelk}  \&
  {Tscharnuter}}{{Morfill} et~al.}{1978}]{Morfill1978}
{Morfill} G.,  {Roeser} S.,  {Voelk} H.,   {Tscharnuter} W.,  1978, \mn@doi
  [Moon and Planets] {10.1007/BF00896994}, \href
  {https://ui.adsabs.harvard.edu/abs/1978M&P....19..211M} {19, 211}

\bibitem[\protect\citeauthoryear{{Nakagawa}, {Nakazawa}  \&
  {Hayashi}}{{Nakagawa} et~al.}{1981}]{Nakagawa1981}
{Nakagawa} Y.,  {Nakazawa} K.,   {Hayashi} C.,  1981, \mn@doi [\icarus]
  {10.1016/0019-1035(81)90018-X}, \href
  {https://ui.adsabs.harvard.edu/abs/1981Icar...45..517N} {45, 517}

\bibitem[\protect\citeauthoryear{{Nakagawa}, {Sekiya}  \& {Hayashi}}{{Nakagawa}
  et~al.}{1986}]{Nakagawa1986}
{Nakagawa} Y.,  {Sekiya} M.,   {Hayashi} C.,  1986, \mn@doi [\icarus]
  {10.1016/0019-1035(86)90121-1}, \href
  {https://ui.adsabs.harvard.edu/abs/1986Icar...67..375N} {67, 375}

\bibitem[\protect\citeauthoryear{{Ohashi} et~al.,}{{Ohashi}
  et~al.}{2021}]{Ohashi2021}
{Ohashi} S.,  et~al., 2021, \mn@doi [\apj] {10.3847/1538-4357/abd0fa}, \href
  {https://ui.adsabs.harvard.edu/abs/2021ApJ...907...80O} {907, 80}

\bibitem[\protect\citeauthoryear{{Ohashi}, {Kobayashi}, {Sai}  \&
  {Sakai}}{{Ohashi} et~al.}{2022}]{Ohashi2022}
{Ohashi} S.,  {Kobayashi} H.,  {Sai} J.,   {Sakai} N.,  2022, arXiv e-prints,
  \href {https://ui.adsabs.harvard.edu/abs/2022arXiv220607800O} {p.
  arXiv:2206.07800}

\bibitem[\protect\citeauthoryear{{Ormel} \& {Cuzzi}}{{Ormel} \&
  {Cuzzi}}{2007}]{OrmelCuzzi2007}
{Ormel} C.~W.,  {Cuzzi} J.~N.,  2007, \mn@doi [\aap]
  {10.1051/0004-6361:20066899}, \href
  {https://ui.adsabs.harvard.edu/abs/2007A&A...466..413O} {466, 413}

\bibitem[\protect\citeauthoryear{{Ormel}, {Paszun}, {Dominik}  \&
  {Tielens}}{{Ormel} et~al.}{2009}]{Ormel2009}
{Ormel} C.~W.,  {Paszun} D.,  {Dominik} C.,   {Tielens} A.~G.~G.~M.,  2009,
  \mn@doi [\aap] {10.1051/0004-6361/200811158}, \href
  {https://ui.adsabs.harvard.edu/abs/2009A&A...502..845O} {502, 845}

\bibitem[\protect\citeauthoryear{{Segura-Cox} et~al.,}{{Segura-Cox}
  et~al.}{2020}]{Segura-Cox2020}
{Segura-Cox} D.~M.,  et~al., 2020, \mn@doi [\nat] {10.1038/s41586-020-2779-6},
  \href {https://ui.adsabs.harvard.edu/abs/2020Natur.586..228S} {586, 228}

\bibitem[\protect\citeauthoryear{{Sheehan} \& {Eisner}}{{Sheehan} \&
  {Eisner}}{2018}]{Sheehan2018}
{Sheehan} P.~D.,  {Eisner} J.~A.,  2018, \mn@doi [\apj]
  {10.3847/1538-4357/aaae65}, \href
  {https://ui.adsabs.harvard.edu/abs/2018ApJ...857...18S} {857, 18}

\bibitem[\protect\citeauthoryear{{Sheehan}, {Tobin}, {Federman}, {Megeath}  \&
  {Looney}}{{Sheehan} et~al.}{2020}]{Sheehan2020}
{Sheehan} P.~D.,  {Tobin} J.~J.,  {Federman} S.,  {Megeath} S.~T.,   {Looney}
  L.~W.,  2020, \mn@doi [\apj] {10.3847/1538-4357/abbad5}, \href
  {https://ui.adsabs.harvard.edu/abs/2020ApJ...902..141S} {902, 141}

\bibitem[\protect\citeauthoryear{{Shu}}{{Shu}}{1977}]{Shu1977}
{Shu} F.~H.,  1977, \mn@doi [\apj] {10.1086/155274}, \href
  {https://ui.adsabs.harvard.edu/abs/1977ApJ...214..488S} {214, 488}

\bibitem[\protect\citeauthoryear{{Smoluchowski}}{{Smoluchowski}}{1916}]{Smoluchowski1916}
{Smoluchowski} M.~V.,  1916, Zeitschrift fur Physik, \href
  {https://ui.adsabs.harvard.edu/abs/1916ZPhy...17..557S} {17, 557}

\bibitem[\protect\citeauthoryear{{Steinacker}, {Pagani}, {Bacmann}  \&
  {Guieu}}{{Steinacker} et~al.}{2010}]{Steinacker_et_al.2010}
{Steinacker} J.,  {Pagani} L.,  {Bacmann} A.,   {Guieu} S.,  2010, \mn@doi
  [\aap] {10.1051/0004-6361/200912835}, \href
  {https://ui.adsabs.harvard.edu/abs/2010A&A...511A...9S} {511, A9}

\bibitem[\protect\citeauthoryear{{Stone}, {Tomida}, {White}  \&
  {Felker}}{{Stone} et~al.}{2020}]{Stone2020}
{Stone} J.~M.,  {Tomida} K.,  {White} C.~J.,   {Felker} K.~G.,  2020, \mn@doi
  [\apjs] {10.3847/1538-4365/ab929b}, \href
  {https://ui.adsabs.harvard.edu/abs/2020ApJS..249....4S} {249, 4}

\bibitem[\protect\citeauthoryear{{Suttner} \& {Yorke}}{{Suttner} \&
  {Yorke}}{2001}]{Suttner2001}
{Suttner} G.,  {Yorke} H.~W.,  2001, \mn@doi [\apj] {10.1086/320061}, \href
  {https://ui.adsabs.harvard.edu/abs/2001ApJ...551..461S} {551, 461}

\bibitem[\protect\citeauthoryear{{Tanaka}, {Himeno}  \& {Ida}}{{Tanaka}
  et~al.}{2005}]{Tanaka2005}
{Tanaka} H.,  {Himeno} Y.,   {Ida} S.,  2005, \mn@doi [\apj] {10.1086/429658},
  \href {https://ui.adsabs.harvard.edu/abs/2005ApJ...625..414T} {625, 414}

\bibitem[\protect\citeauthoryear{{Testi} et~al.,}{{Testi}
  et~al.}{2014}]{Testi2014review}
{Testi} L.,  et~al., 2014, in {Beuther} H.,  {Klessen} R.~S.,  {Dullemond}
  C.~P.,   {Henning} T.,  eds, Protostars and Planets VI. p.~339 (\mn@eprint
  {arXiv} {1402.1354}), \mn@doi{10.2458/azu\_uapress\_9780816531240-ch015}

\bibitem[\protect\citeauthoryear{Tobin et~al.,}{Tobin et~al.}{2013}]{Tobin2013}
Tobin J.~J.,  et~al., 2013, \mn@doi [The Astrophysical Journal]
  {10.1088/0004-637x/779/2/93}, 779, 93

\bibitem[\protect\citeauthoryear{{Tsukamoto}, {Machida}  \&
  {Inutsuka}}{{Tsukamoto} et~al.}{2021}]{Tsukamoto2021}
{Tsukamoto} Y.,  {Machida} M.~N.,   {Inutsuka} S.-i.,  2021, \mn@doi [\apjl]
  {10.3847/2041-8213/ac2b2f}, \href
  {https://ui.adsabs.harvard.edu/abs/2021ApJ...920L..35T} {920, L35}

\bibitem[\protect\citeauthoryear{{Voelk}, {Jones}, {Morfill}  \&
  {Roeser}}{{Voelk} et~al.}{1980}]{Voelk1980}
{Voelk} H.~J.,  {Jones} F.~C.,  {Morfill} G.~E.,   {Roeser} S.,  1980, \aap,
  \href {https://ui.adsabs.harvard.edu/abs/1980A&A....85..316V} {85, 316}

\bibitem[\protect\citeauthoryear{{Wong}, {Hirashita}  \& {Li}}{{Wong}
  et~al.}{2016}]{Wong2016}
{Wong} Y. H.~V.,  {Hirashita} H.,   {Li} Z.-Y.,  2016, \mn@doi [\pasj]
  {10.1093/pasj/psw066}, \href
  {https://ui.adsabs.harvard.edu/abs/2016PASJ...68...67W} {68, 67}

\bibitem[\protect\citeauthoryear{{Yen}, {Koch}, {Takakuwa}, {Krasnopolsky},
  {Ohashi}  \& {Aso}}{{Yen} et~al.}{2017}]{Yen2017}
{Yen} H.-W.,  {Koch} P.~M.,  {Takakuwa} S.,  {Krasnopolsky} R.,  {Ohashi} N.,
  {Aso} Y.,  2017, \mn@doi [\apj] {10.3847/1538-4357/834/2/178}, \href
  {https://ui.adsabs.harvard.edu/abs/2017ApJ...834..178Y} {834, 178}

\bibitem[\protect\citeauthoryear{Youdin \& Goodman}{Youdin \&
  Goodman}{2005}]{YoudinGoodman2005}
Youdin A.~N.,  Goodman J.,  2005, \mn@doi [The Astrophysical Journal]
  {10.1086/426895}, 620, 459

\bibitem[\protect\citeauthoryear{Zhang, Bergin, Blake, Cleeves, Hogerheijde,
  Salinas  \& Schwarz}{Zhang et~al.}{2016}]{Zhang_2016}
Zhang K.,  Bergin E.~A.,  Blake G.~A.,  Cleeves L.~I.,  Hogerheijde M.,
  Salinas V.,   Schwarz K.~R.,  2016, \mn@doi [The Astrophysical Journal]
  {10.3847/2041-8205/818/1/l16}, 818, L16

\makeatother
\end{thebibliography}




\appendix

\section{Brownian motion}

\begin{figure}
    \centering
    \includegraphics[width=\columnwidth]{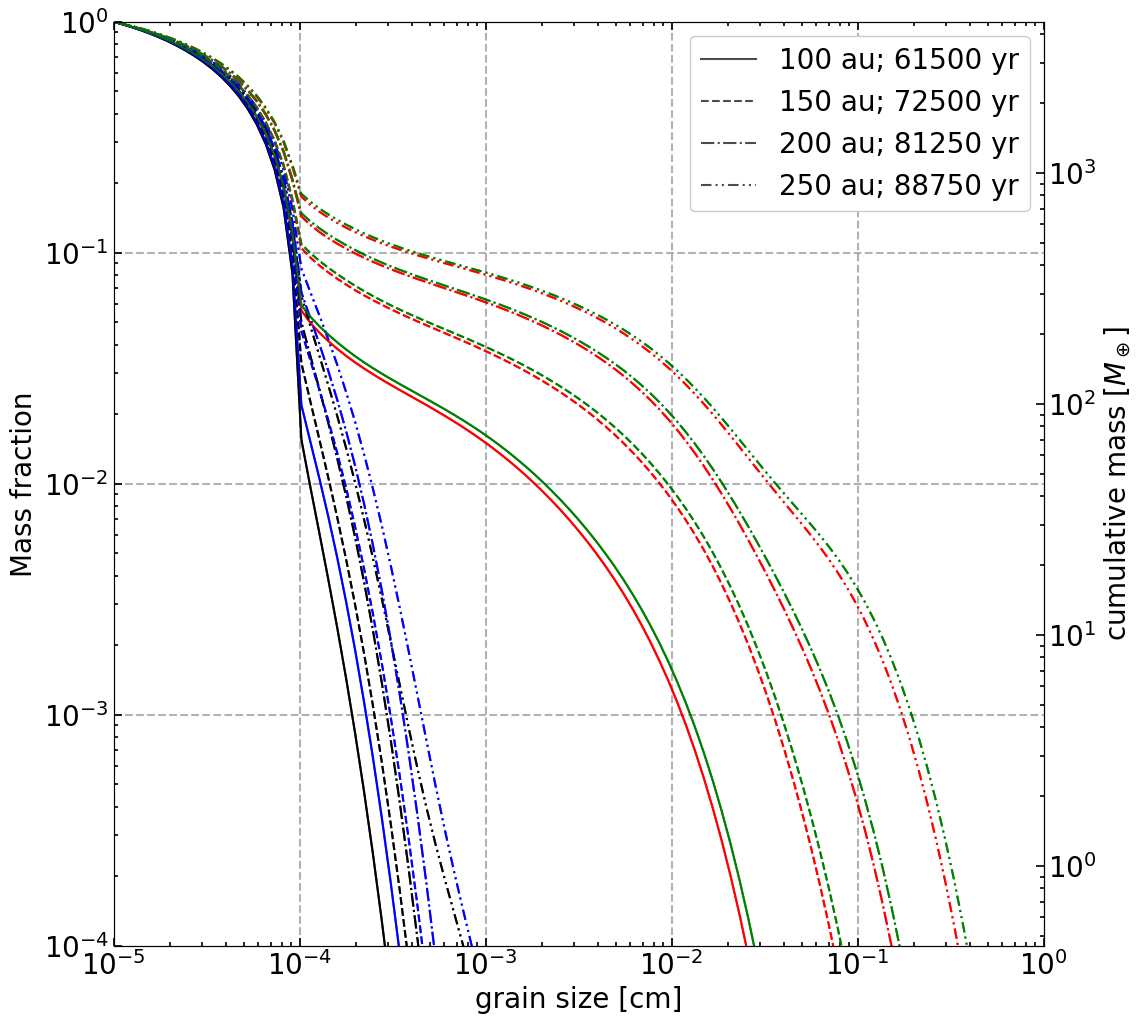}
    \caption{Cumulative grain mass as a function of grain size similar to Fig.~\ref{fig:74_34-74_32_summary_plot} but with the Brownian motion included. The mass includes both the grains in the active computational domain and those advected through the inner boundary. The black and red lines are the baseline models for $\chi=1$ and $4$ respectively without Brownian motion; the blue and green lines are the corresponding models with Brownian motion included. }
    \label{fig:74_39-78_addBM_summary}
\end{figure}

Brownian motion was shown to be important for the growth of small, (sub)$\mu$m-sized grains \citep[e.g.,][]{Suttner2001, Tanaka2005}. In this Appendix, we consider the effects of the Brownian motion on the grain growth in the two simulations shown in Fig.~\ref{fig:74_34-74_32_summary_plot}. The relative speed induced by Brownian motion on two dust particles of mass $m_1$ and $m_2$ is given by
\begin{equation}
    \Delta v_\mathrm{BM} = \sqrt{\frac{8k_bT}{\pi}\frac{m_1 + m_2}{m_1m_2}}.
\end{equation}
It is added to the relative speed induced by the dust-gas drift in grain coagulation calculation. 
The results are shown in Fig.~\ref{fig:74_39-78_addBM_summary}, where the cumulative mass of the grain up to a certain grain size is plotted as a function of the grain size for both the reference run without any enhanced collision speed ($\chi=1$) and with an enhancement factor of $\chi=4$ and with and without the Brownian motion. It is clear that the Brownian motion increases the grain sizes at a given time somewhat, but the increase is relatively modest. This is to be expected because the later grain growth to larger sizes is relatively insensitive to the size distribution of the smallest grains that are most affected by the Brownian motion. We conclude that the inclusion of the Brownian motion cannot explain the abundance of large mm/cm sized grains that are inferred in young protostellar disks and additional pieces of physics are needed.

%

\bsp    
\label{lastpage}
\end{document}